\documentclass[aps, prd, twocolumn, superscriptaddress,tightenlines, preprintnumbers,nofootnoteinbib]{revtex4-1}
\usepackage[T1]{fontenc}
\usepackage{microtype}
\usepackage[textsize=scriptsize,backgroundcolor=red!70,linecolor=red]{todonotes}
\usepackage{booktabs}
\usepackage{dcolumn}
\usepackage{amsmath}
\usepackage{amsfonts}
\usepackage{amssymb}
\usepackage{graphicx}
\usepackage{ulem}
\usepackage{hyperref}
\usepackage[all]{hypcap}
\usepackage{paralist}
\usepackage{multirow}
\usepackage{mathrsfs}
\usepackage{graphicx}
\usepackage{dcolumn}
\usepackage{bm}
\usepackage{epsf}
\usepackage{dcolumn}

\usepackage{footmisc}
\usepackage{slashed}
\usepackage{xcolor}
\usepackage{float}

\begin{document}

\setlength{\abovedisplayskip}{5pt}
\setlength{\belowdisplayskip}{5pt}
\setlength{\abovedisplayshortskip}{5pt}
\setlength{\belowdisplayshortskip}{5pt}

\preprint{}

\title{Impact of coherent scattering on relic neutrinos boosted by cosmic rays}

\author{Jiajie Zhang}
\email{zhangjj253@mail2.sysu.edu.cn}
\affiliation{School of Physics, Sun Yat-sen University, Guangzhou, 510275, China}

\author{Alexander Sandrock}
\email{asandrock@uni-wuppertal.de}
\affiliation{Faculty of Mathematics and Natural Sciences, University of Wuppertal, 42119 Wuppertal, Germany}

\author{Jiajun Liao}
\email{liaojiajun@mail.sysu.edu.cn}
\affiliation{School of Physics, Sun Yat-sen University, Guangzhou, 510275, China}

\author{Baobiao Yue}
\email{bayue@uni-wuppertal.de}
\affiliation{Faculty of Mathematics and Natural Sciences, University of Wuppertal, 42119 Wuppertal, Germany}

\begin{abstract}
Ultra-high-energy cosmic rays (UHECR) scattering off the cosmic relic neutrino background have recently gained renewed interest in the literature. Current data suggest that (UHECR) are predominantly made of heavy nuclei. 
Similar to the coherent elastic neutrino-nucleus scattering (CE$\nu$NS) observed at low-energy neutrino experiments, the cross section of heavy nucleus scattering off relic neutrinos will be coherently enhanced since the energy of relic neutrinos can reach $\sim\mathcal{O}(10)$ MeV in the rest frame of the UHECR.
We calculate the diffuse flux of relic neutrinos boosted by UHECR after taking into account the contributions from both coherent and incoherent scatterings. Using current data from IceCube and Pierre Auger Observatory, we place constraints on the overdensity of relic neutrinos down to $\sim 10^8$. Since the flux of boosted relic neutrinos peaks at an energy of $\sim 200\, \text{PeV}$, we also entertain the possibility to explain the recently observed KM3NeT event with boosted relic neutrinos from UHECR.
\end{abstract}

\maketitle

\section{Introduction}

The cosmic relic neutrino background (C$\nu$B), a fundamental prediction of the standard cosmological model ($\Lambda$CDM), provides a unique probe of the early Universe and the evolution of cosmic structure~\cite{TopicalConvenersKNAbazajianJECarlstromATLee:2013bxd,Giunti:2007ry,Lesgourgues:2013sjj}. The $\Lambda$CDM model predicts that the C$\nu$B today has an average number density of  $56 \, \text{cm}^{-3}$ per spin and flavor degree of freedom with a temperature of $1.95 \, \text{K}$~\cite{Giunti:2007ry,Lesgourgues:2013sjj}. 
Detecting these relic neutrinos is quite challenging, and requires experiments with ultralow energy thresholds. The proposed PTOLEMY experiment~\cite{PTOLEMY:2018jst}, which employs the method of capturing relic neutrinos in tritium, is hampered by the Heisenberg uncertainty principle for storing tritium atoms on a graphene sheet~\cite{PTOLEMY:2022ldz}. The neutrino number density can be greatly enhanced in a nonstandard cosmological history ~\cite{Bondarenko:2023ukx}, such as those involving dark matter decay into neutrinos~\cite{McKeen:2018xyz, Chacko:2018uke, Nikolic:2020fom}, or new Yukawa interactions of neutrinos~\cite{Smirnov:2022sfo}. The current strongest experimental constraint is set by the KATRIN experiment, with an upper limit on the local C$\nu$B overdensity $\eta < 9.7 \times 10^{10}$ at the 90\% confidence level (CL)~\cite{KATRIN:2022kkv}.

Ultra-high-energy (UHE) cosmic rays (CR) scattering off relic neutrinos provide a novel approach to probe the existence of C$\nu$B, a concept first introduced by Hara and Sato in the 1980s~\cite{Hara:1979he,Hara:1980mz}. Recent studies have used  CR scattering off the C$\nu$B to impose constraints on the C$\nu$B overdensity in various contexts. Ref.~\cite{Ciscar-Monsalvatje:2024tvm} constrains the overdensity in the ballpark of $\sim 10^{13}$ within the Milky Way and $\sim 10^{11}$ near the blazar TXS 0506+056. A very strong limit of $\sim 10^4$ on the average C$\nu$B overdensity has been placed in ~Ref.~\cite{Herrera:2024upj}. By considering both elastic scattering (ES) and deep inelastic scattering (DIS), Ref.~\cite{DeMarchi:2024zer} put a constraint on the weighted overdensities in cosmic-ray reservoirs down to $\sim 10^{10}$. Note that Refs.~\cite{Ciscar-Monsalvatje:2024tvm,Herrera:2024upj} only consider the proton composition in CR with an oversimplified scattering cross sections, neglecting the significant contribution from heavy nuclei to boost the C$\nu$B.

Current data indicate a dominant heavy-element composition for UHECR, with the proton fraction dropping below 10\% for CR energy above $10 \, \text{EeV}$~\cite{PierreAuger:2022atd}, which is also consistent with air shower simulations despite of large uncertainties in hadronic interactions~\cite{Ehlert:2023btz}. A pure iron model for UHECR composition is even adopted in some studies~\cite{Armengaud:2004yt, Allard:2008gj,Arisaka:2007iz}. In addition, observations from the Pierre Auger Observatory~\cite{PierreAuger:2022atd}, KASCADE-Grande~\cite{Apel:2013uni}, and Telescope Array~\cite{TelescopeArray:2018bep} also show an increasing prevalence of heavy nuclei at the highest energies. Given the significant presence of heavy nuclei in UHECR, the contributions from heavy nuclei in boosting the C$\nu$B cannot be ignored.

For neutrinos with an energy of $\sim\mathcal{O}(10) \, \text{MeV}$, coherent elastic neutrino-nucleus scattering (CE$\nu$NS) occurs since neutrinos interact with all nucleons in a nucleus collectively, which significantly increases the cross section at low energy neutrino scattering experiments~\cite{Freedman:1973yd}. CE$\nu$NS has been confirmed experimentally by the COHERENT Collaboration in 2017 using accelerator-based sources~\cite{COHERENT:2017ipa}. Subsequent measurements at COHERENT further validated the existence of  CE$\nu$NS~\cite{COHERENT:2020iec,COHERENT:2024axu}, complemented by observations from the solar~\cite{XENON:2024ijk,PandaX:2024muv} and reactor data~\cite{Ackermann:2025obx}. In the context of an iron nucleus with an energy $\sim 10 \, \text{EeV}$ scattering on a relic neutrino of mass $0.1 \, \text{eV}$, the neutrino energy in the rest frame of the nucleus is $\sim 20 \, \text{MeV}$, which is within the same energy range in which CE$\nu$NS becomes dominant at low energy neutrino scattering experiments. Thus, the coherent enhancement of the cross section for heavy nuclei in scattering off the relic neutrinos can be significant.

In this paper, we study the diffuse flux of the C$\nu$B boosted by UHECR, taking into account the contributions from both the coherent and incoherent scattering channels. 
In Sec.~II, we derive the coherent and incoherent scattering cross sections for UHECR colliding with nonrelativistic C$\nu$B.
In Sec.~III, we obtain the UHECR flux and calculate the boosted C$\nu$B flux. In Sec.~IV, we use current data from IceCube (IC) and the Pierre Auger Observatory (PAO) to derive constraints on the C$\nu$B overdensity.
We summarize our results in Sec.~V.
\begin{figure}[t!]
    \centering
    \includegraphics[width=0.45\textwidth, trim=80 160 90 170, clip]{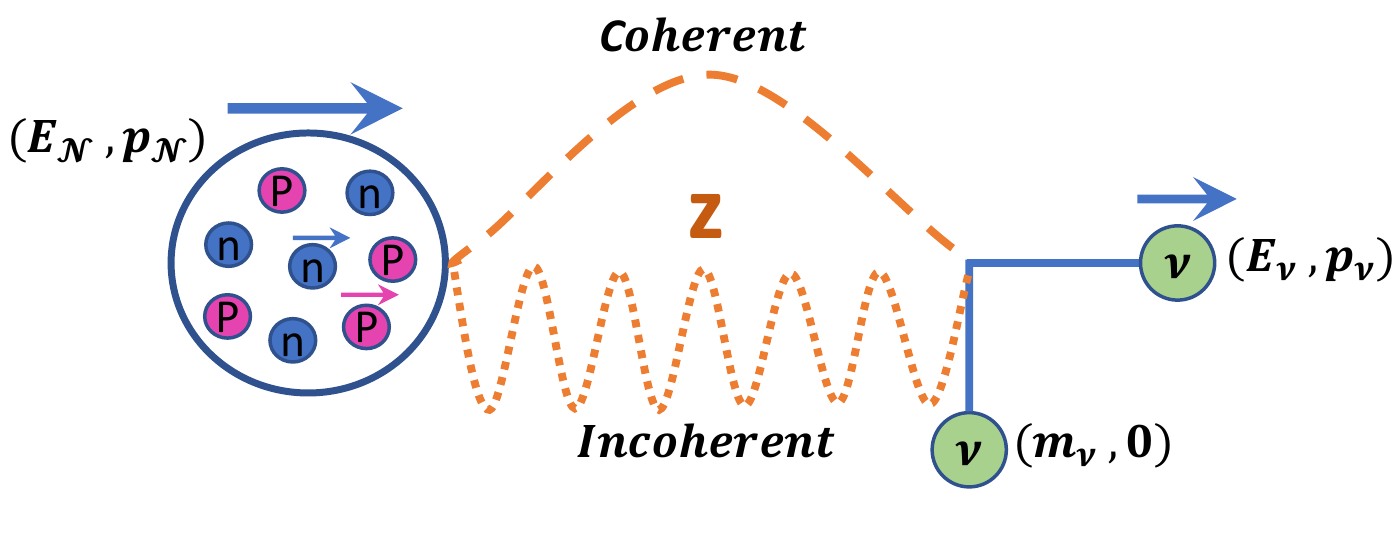}
    \caption{Schematic diagram showing the C$\nu$B boosted by UHECR. A nucleus $\mathcal{N}$ with energy $E_{\mathcal{N}}$ scatters off a relic neutrino, boosting its energy to $E_\nu$. At a small momentum transfer, the wavelength of the $Z$ boson is comparable to the nucleus, and coherent scattering (dashed line) dominates. At a large momentum transfer, the individual nucleons inside a nucleus are resolved by the $Z$ boson, and incoherent scattering (dotted line) becomes important.}
    \label{scattering_schematic}
\end{figure}

\section{Cross Sections}
Here we only consider the lightest neutrino mass $m_\text{1ightest} \gtrsim 10^{-3}\,\text{eV}$ (which is larger than the C$\nu$B temperature $ T_\nu \simeq 10^{-4}\,\text{eV}$) for simplicity.  Thus, the C$\nu$B can be treated approximately as at rest in the laboratory frame.
As shown in Fig.~\ref{scattering_schematic}, UHECR nuclei scattering off the C$\nu$B consist of two channels: the coherent elastic neutrino-nucleus scattering and incoherent scattering~\cite{Bednyakov:2018mjd,Bednyakov:2021lul}. Similar to the dark matter-nucleus scattering in Refs.~\cite{Bednyakov:2023pwi, BetancourtKamenetskaia:2025noa, Ge:2024euk}, the total differential cross section can be written as the sum of the two contributions~\cite{Bednyakov:2018mjd,Bednyakov:2021lul}, i.e.,
\begin{equation}
\label{eq:total_sum}
\frac{d \sigma^{\nu \mathcal{N}_i}}{d E_\nu} = \frac{d \sigma^{\nu \mathcal{N}_i}_\text{coh}}{d E_\nu} + \frac{d \sigma^{\nu \mathcal{N}_i}_\text{incoh}}{d E_\nu}\,,
\end{equation}
where $E_\nu$ is the energy of the boosted C$\nu$B.
At a small momentum transfer, in which the wavelength of $Z$ boson is comparable to the nuclear radius, the relic neutrino interacts coherently with the entire nucleus, and the differential cross section in the laboratory frame is
\begin{equation}
\frac{d \sigma^{\nu \mathcal{N}_i}_\text{coh}}{d E_\nu} = \frac{G_F^2 m_\nu}{\pi}\, Q_{W,i}^2 \left( 1 - \frac{E_\nu}{E_{\mathcal{N}_i}} - \frac{m_{\mathcal{N}_i}^2 E_\nu}{2 m_\nu E_{\mathcal{N}_i}^2} \right) F^2(q^2)\,,
\label{eq:Coherent}
\end{equation}
where $m_{\mathcal{N}_i}$ ($E_{\mathcal{N}_i}$) is the mass (energy) of the nucleus $\mathcal{N}_i$, $m_\nu$ is the neutrino mass, $q = \sqrt{2 m_\nu E_\nu}$ is the momentum transfer, $F(q^2)$ is the nuclear form factor~\cite{Klein:1999qj},
$Q_{W,i} = Z_i g_V^p + N_i g_V^n$ is the nuclear weak charge with $Z_i$ ($N_i$) the proton (neutron) numbers of $\mathcal{N}_i$, and $g_V^{p,n}$ the vector coupling constants. 
Note that we assume a lepton–symmetric cosmology with a equal number for neutrinos and antineutrinos~\cite{Grohs:2020xxd,Long:2014zva}. We summed the cross sections over neutrinos and antineutrinos for simplicity; i.e. we take $d \sigma^{\nu \mathcal{N}_i} / d E_\nu \equiv (d \sigma^{\nu \mathcal{N}_i} / d E_\nu + d \sigma^{\bar{\nu} \mathcal{N}_i} / d E_\nu)$.
In Appendix A, we derive the coherent scattering cross sections for both Dirac and Majorana neutrinos, and we find that the difference between them is negligible.

At a high momentum transfer, the $Z$ boson resolves individual nucleons within the nucleus, which leads to incoherent scattering. The cross section can be approximated as a linear superposition of neutrino-nucleon scatterings. In the laboratory frame, the incoherent cross section is
\begin{align}
\label{eq:incoherent}
\frac{d \sigma^{\nu \mathcal{N}_i}_\text{incoh}}{d E_\nu} = \left[ Z_i \frac{d \sigma_{\text{ES}}^{\nu p}}{d E_\nu} + N_i \frac{d \sigma_{\text{ES}}^{\nu n}}{d E_\nu} \right] (1 - F^2(q^2))\,.
\end{align}
Here, the differential cross section of neutrinos scattering on a nucleon (denoted $N$) within the heavy nuclei is taken from the standard elastic scattering  process~\cite{Giunti:2007ry,DeMarchi:2024zer,Formaggio:2012cpf},
\begin{equation}
\frac{d \sigma_{\text{ES}}^{\nu N}}{d E_\nu} = \frac{G_{\text{F}}^2 m_\nu m_N^4}{\pi (s - m_N^2)^2} \left[ A_N(q^2) + C_N(q^2) \frac{(s - u)^2}{m_N^4} \right]\,,
\label{eq:ES-cross-section}
\end{equation}
where the Mandelstam variables $s \approx 2 m_\nu E_N + m_N^2$, $t \approx -q^2$, and $u \approx m_N^2 - 2 m_\nu (E_N - E_\nu)$ with $m_N$ ($E_N$) being the nucleon mass (energy). The coefficients $A_N(q^2)$ and $C_N(q^2)$ depend on the momentum transfer $q$ and the nucleon form factors. 
We have checked that Eq.~\eqref{eq:incoherent} is consistent with Eq.~(46) in Ref.~\cite{Bednyakov:2018mjd} for unpolarized nuclei; see Appendix A for more details.
\begin{figure}[t!]
    \centering
    \includegraphics[width=0.45\textwidth]{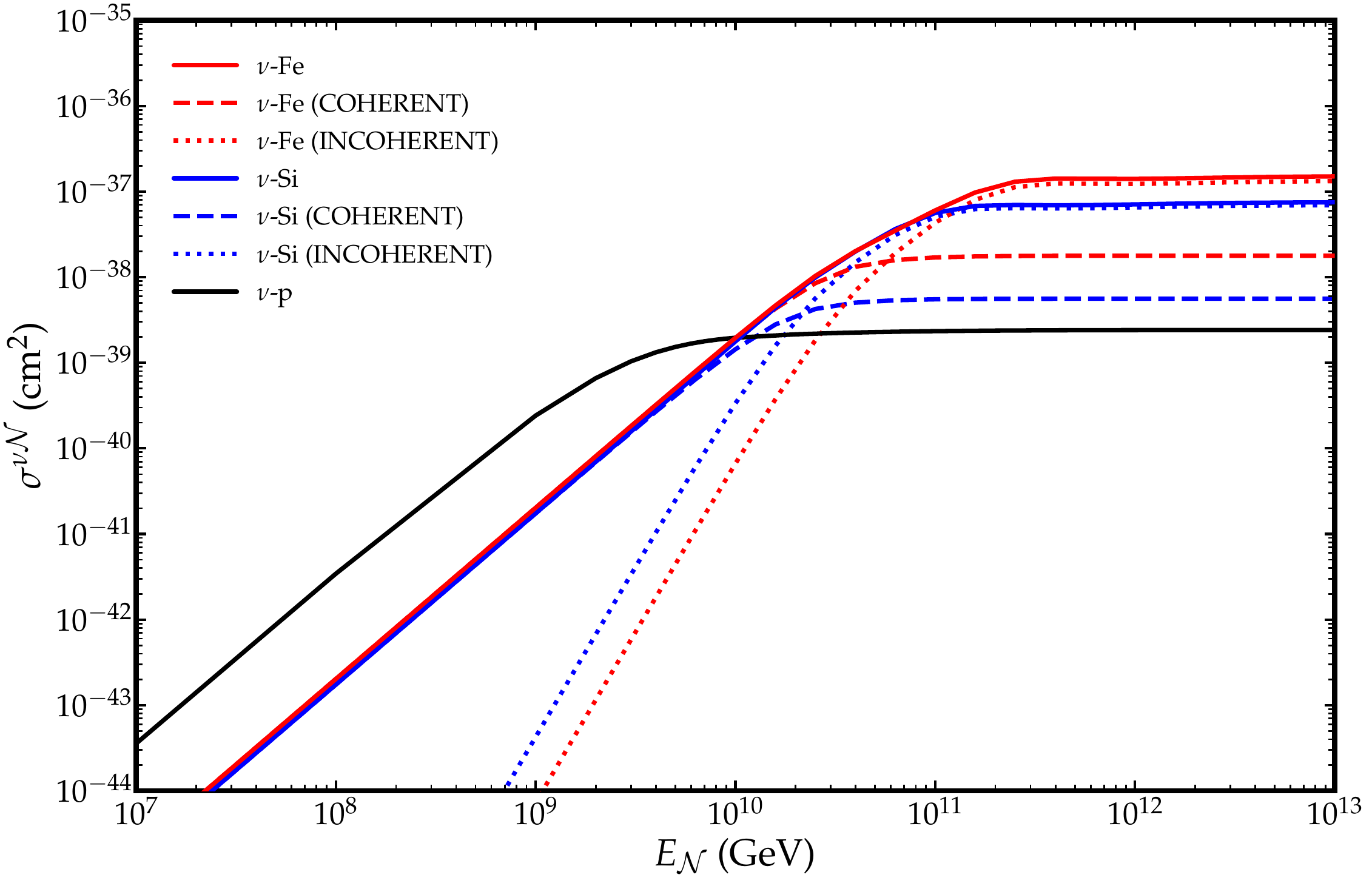}
    \includegraphics[width=0.45\textwidth]{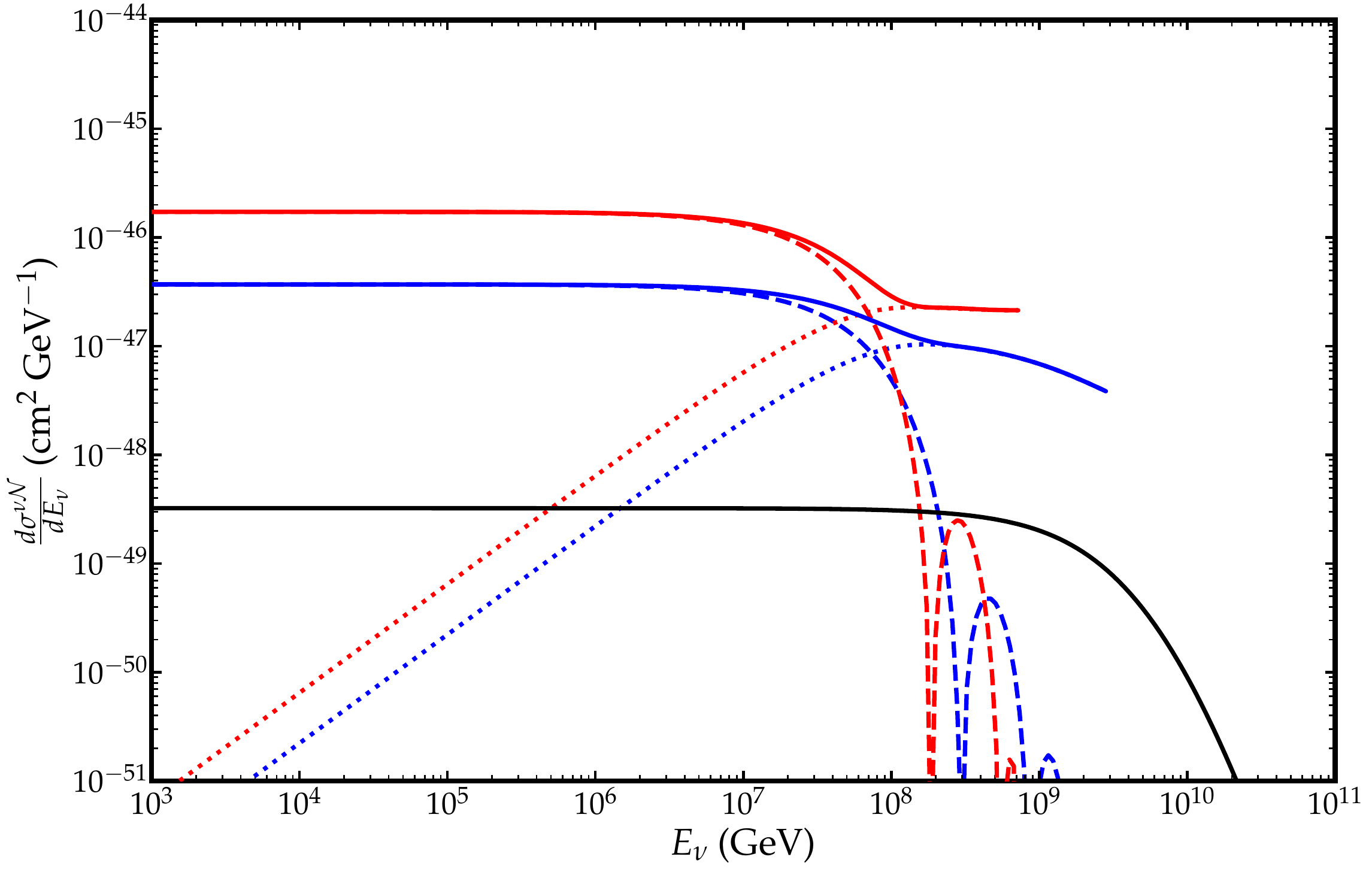}
    \caption{Upper panel: The total neutrino-nucleus scattering cross section as a function of the UHECR energy $E_{\mathcal{N}}$. Lower panel: The differential scattering cross section as a function of the boosted neutrino energy $E_\nu$ with $E_{\mathcal{N}} = 100 \, \text{EeV}$. In both panels, red, blue, and gray solid lines correspond to the C$\nu$B boosted by the iron, silicon, and proton, respectively. The dashed (dotted) lines denote the coherent (incoherent) contributions. Here, we fix $m_\nu = 0.1 \, \text{eV}$.}
    \label{fig:cross_section}
\end{figure}

To illustrate the coherent and incoherent contributions to the cross section of different UHECR nuclei, we show the total cross section as a function of the UHECR energy $E_{\mathcal{N}}$ for proton (p), silicon (Si), and iron (Fe) in the upper panel of Fig.~\ref{fig:cross_section} for $m_\nu = 0.1 \, \text{eV}$. We see that for $E_{\mathcal{N}} \lesssim 10 \, \text{EeV}$, the proton elastic scattering cross section dominates, while for $E_{\mathcal{N}}$ above $\sim$10 EeV, heavier nuclei Si and Fe exhibit larger cross sections. As $E_{\mathcal{N}}$ increases, the cross section curve reaches a plateau because the coefficients $A_N(q^2)$ and $C_N(q^2)$ approach zero at large momentum transfer $q$ due to the suppression of the nucleon form factors. For a heavier nucleus with a larger mass number,  the suppression of the nucleon form factor occurs at a higher nucleus energy since each nucleon carrying a fraction $1/A$ of the nucleus energy $E_{\mathcal{N}}$. The lower panel of Fig.~\ref{fig:cross_section} shows the differential cross section as a function of boosted neutrino energy $E_\nu$ for $E_{\mathcal{N}} = 100 \, \text{EeV}$ . Since $q = \sqrt{2 m_\nu E_\nu}$, from Eq.~\eqref{eq:Coherent}, we see that coherent scattering dominates at small $E_\nu$, where $F^2(q^2) \approx 1$ and $d\sigma_{\text{coh}}/dE_\nu \propto N_i^2$. From Eq.~\eqref{eq:incoherent}, we see that incoherent scattering dominates at large $E_\nu$, where $F^2(q^2) \approx 0$ and $d\sigma_{\text{incoh}} /dE_\nu\propto A_i$. For $E_\nu \lesssim 10^8 \, \text{GeV}$, the differential cross section of Fe is approximately 2 orders of magnitude larger than the proton elastic scattering cross section, reflecting the coherent enhancement in the neutrino-nucleus scattering. Furthermore, there is a cutoff for $E_\nu$ for a given $E_{\mathcal{N}}$, and the cutoff for heavier nuclei occurs at a lower $E_\nu$, since the maximum of boosted neutrino energy $E_\nu^{\text{max}} = E_{\mathcal{N}_i}^2 / (E_{\mathcal{N}_i} + m_{\mathcal{N}_i}^2 / 2 m_\nu)$ decreases as the nuclear mass $m_{\mathcal{N}_i}$ increases.

\section{Boosted C$\nu$B flux}

UHECR travel vast distances from their sources to the Earth and scatter with the C$\nu$B along their paths, leading to a diffuse flux of boosted neutrinos at very high energies. 
The differential flux of the boosted C$\nu$B at Earth can be written as
\begin{multline}
\frac{d \phi_{\nu}}{d E_{\nu}} 
= \sum_{i,j} \int_{z_{\text{min}}}^{z_{\text{max}}} dz \frac{c}{H(z)} \eta n_{\nu_j} (1+z)^3  \\
\times \int_{0}^{\infty} d E_{\mathcal{N}_i} 
\, \frac{d \sigma^{\nu \mathcal{N}_i}}{d E'_\nu}\,\frac{d\phi_{\mathcal{N}_i}}{dE_{\mathcal{N}_i}}(z)\,
\Theta \!\left[ E_{\nu}^{\max} - E'_{\nu} \right],
\label{eq:flux}
\end{multline}
where $i$ and $j$ label the CR nucleus species and the neutrino mass eigenstate, respectively. 
Here, $n_{\nu_j}$ is the present C$\nu$B number density, $\eta$ is the overdensity factor, and $d\phi_{\mathcal{N}_i}/dE_{\mathcal{N}_i}(z)$ is the UHECR flux at redshift $z$. $H(z) = H_0 \sqrt{\Omega_m(1+z)^3 + \Omega_\Lambda}$ is the Hubble expansion rate, with $c/H_0 = 1.1267\times10^{28}\,\text{cm}$, $\Omega_m = 0.315$, and $\Omega_\Lambda = 0.685$~\cite{ParticleDataGroup:2024cfk}. 
The redshift integral spans from $z_{\min}=0$ to $z_{\max}=6$~\cite{Hopkins:2006bw,Herrera:2024upj}.

The differential cross section $d \sigma^{\nu \mathcal{N}_i}/d E'_\nu$ depends on the CR energy $E_{\mathcal{N}_i}$ and the neutrino energy $E'_\nu = E_\nu (1+z)$ at redshift $z$. Note that our Eq.~\eqref{eq:flux} differs from Eq.(1) in Ref. \cite{Herrera:2024upj}, which uses $\sigma / E_\nu^{\max}$ as an approximation for the differential neutrino-proton scattering cross section $d \sigma^{\nu p}/d E_\nu$. This approach agrees with our results at low energies. However, at high energies, $d \sigma^{\nu p}/d E_\nu$ is suppressed by the nucleon form factors, whereas this suppression is absent if $\sigma / E_\nu^{\max}$ is used, which will lead to an overestimation of the cross section and boosted neutrino flux; see Appendix C for a detailed comparison of the two cross sections.

In this work, we obtain the UHECR flux $d\phi_{\mathcal{N}_i}/dE_{\mathcal{N}_i}(z)$ using two approaches: (a) one based on simulations with the UHECR propagation code PriNCe~\cite{Heinze:2019jou}; (b) one based on a phenomenological parametrization as in Ref.~\cite{Herrera:2024upj}.
As shown in Appendix B, 
we find that scattering with the C$\nu$B has a negligible effect on UHECR propagation across most of the relevant parameter space. Therefore, we use the UHECR propagation code PriNCe~\cite{Heinze:2019jou} to simulate the UHECR flux at different redshifts, without including C$\nu$B scattering effects.
In our PriNCe simulations, we consider three representative UHECR source evolution models. 
In the main text, we focus on the cosmic star formation rate (SFR) evolution, while the other source evolutions, quasistellar objects (QSO) and gamma-ray bursts (GRB), are presented in Appendix B.
For comparison, we also adopt the mixed composition Hillas parametrization~\cite{Hillas:2005cs, Gaisser:2013bla} to describe the present day CR energy spectrum. 
To ensure consistency with current UHECR observations, we introduce an overall normalization factor $k_\text{best-fit} = 0.601$ for the Hillas model, which is determined by fitting the model to the energy spectrum measured by PAO.
The UHECR flux at higher redshifts is then obtained by rescaling the Hillas spectrum with the CR source distribution function, following the approach used in previous work~\cite{Herrera:2024upj}. 
The details of this procedure are discussed in Appendix B.

In the upper panel of Fig.~\ref{fig:cosmic_ray_flux}, we show the UHECR flux at Earth obtained from the PriNCe simulation with the SFR source distribution, while the UHECR flux obtained from the Hillas model is shown in the lower panel of Fig.~\ref{fig:cosmic_ray_flux}.
The two approaches yields UHECR fluxes that agree with the measured data at PAO~\cite{Veberic:2017hwu}. However, they predict different nuclear composition fractions, and the Hillas model shows a larger fraction of heavy nuclei compared to the PriNCe simulation.
Even though transport codes such as PriNCe can simulate the UHECR flux, our knowledge of extragalactic cosmic rays remains limited ~\cite{PierreAuger:2016use,AlvesBatista:2019tlv}. In particular, different UHECR injection composition fractions remain compatible with current PAO observations~\cite{PierreAuger:2016use}. In earlier studies, such scenarios included a proton-dominated mixed composition, a pure iron composition, or a mixed
composition containing roughly $30\%$ iron, all of which were consistent with the UHECR observations available at that time~\cite{Kotera:2010yn,Allard:2008gj, Allard:2006mv}.

In general, cosmic rays must reach sufficiently high energies to escape magnetic confinement in their host galaxies~\cite{AlvesBatista:2019tlv}. 
Composition and anisotropy measurements indicate that the transition between Galactic and extragalactic cosmic rays occurs near the second knee at $\sim 5\times10^{8}\,\mathrm{GeV}$~\cite{Kachelriess:2019oqu}. 
This is consistent with the results of the PriNCe simulations used in this work. As shown in the upper panel of Fig.~\ref{fig:cosmic_ray_flux},
the PriNCe simulations also indicate that CR above $5\times10^{8}\,\mathrm{GeV}$ are predominantly of extragalactic origin.
Thus, we only consider CR with energy above $5\times10^{8}\,\mathrm{GeV}$, and we have verified that CR with energy below $5\times10^{8}\,\mathrm{GeV}$ contribute negligibly to the boosted C$\nu$B flux. 

\begin{figure}[t!]
    \centering
    \includegraphics[width=0.49\textwidth]{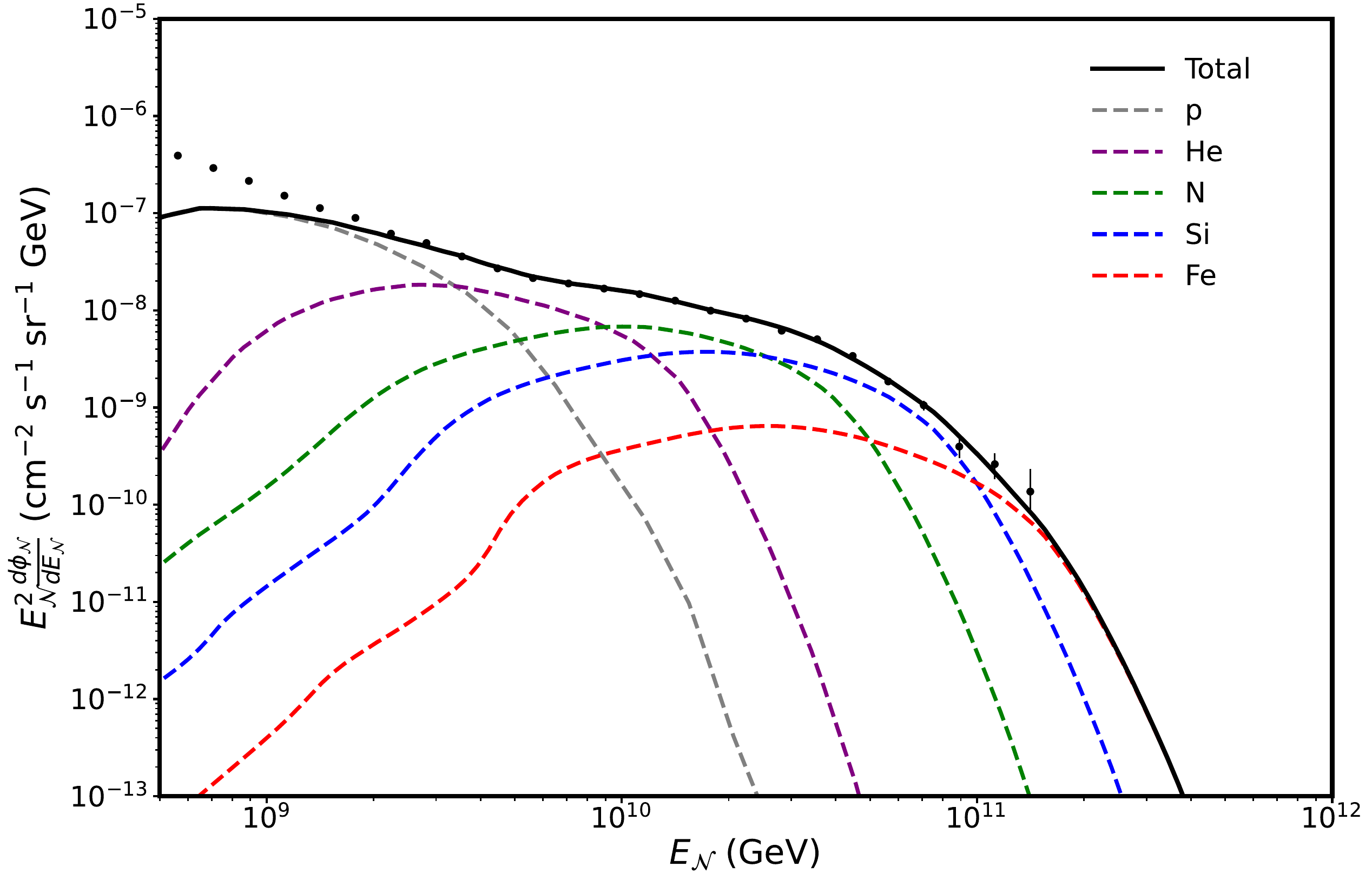}
    \includegraphics[width=0.49\textwidth]{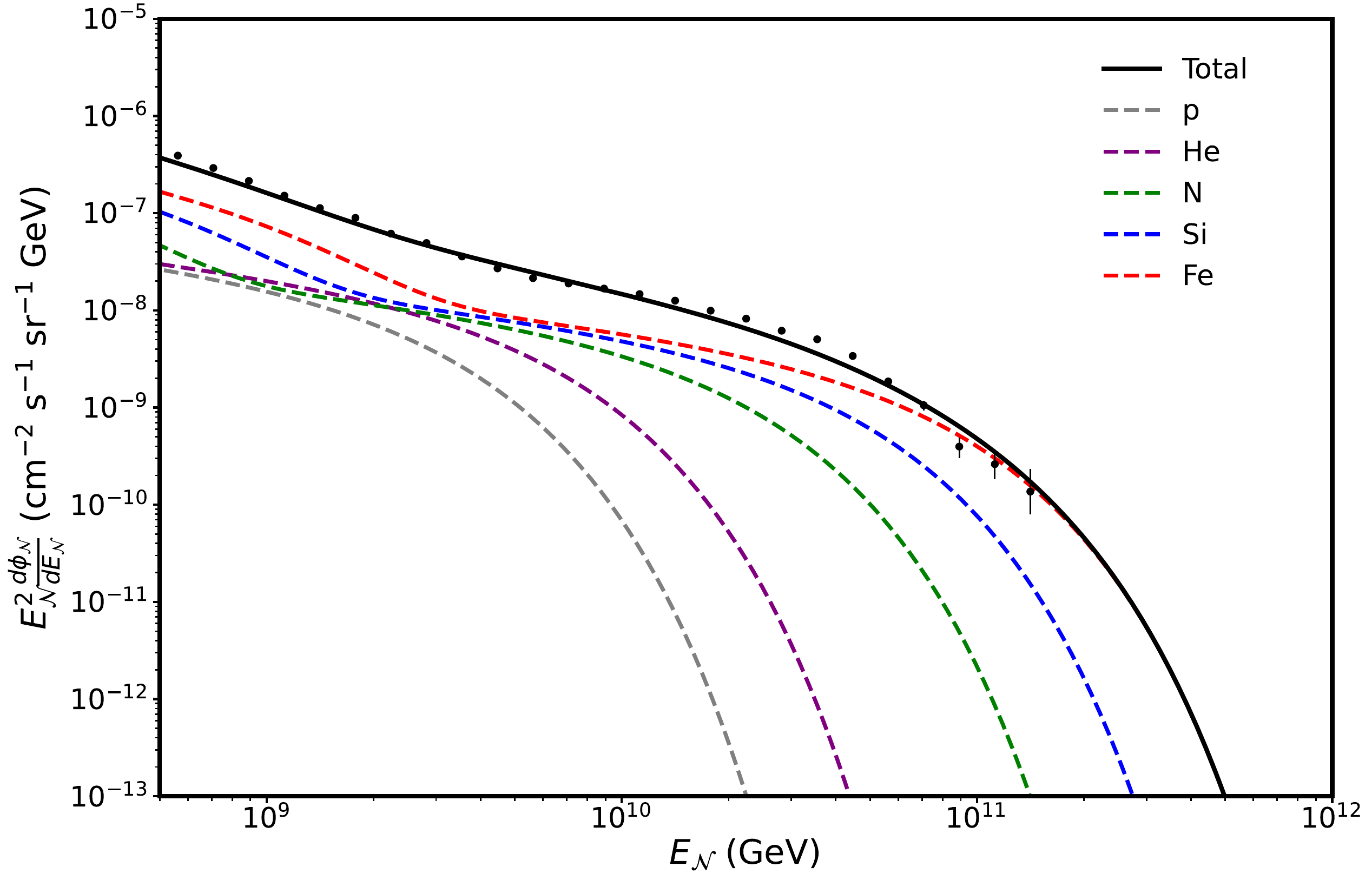}
    \caption{
    Upper panel: The UHECR flux at Earth obtained from the PriNCe simulation with the SFR source distribution.
    Lower panel: The UHECR flux described by the Hillas model normalized by a factor $k_\text{best-fit} = 0.601$.
    Different colors represent flux contributions from different CR elements, while the black solid line shows the total flux.
    The data points are taken from the PAO measurement~\cite{Veberic:2017hwu}.
    }
    \label{fig:cosmic_ray_flux}
\end{figure}

\begin{figure}[t!]
    \centering
    \includegraphics[width=0.49\textwidth]{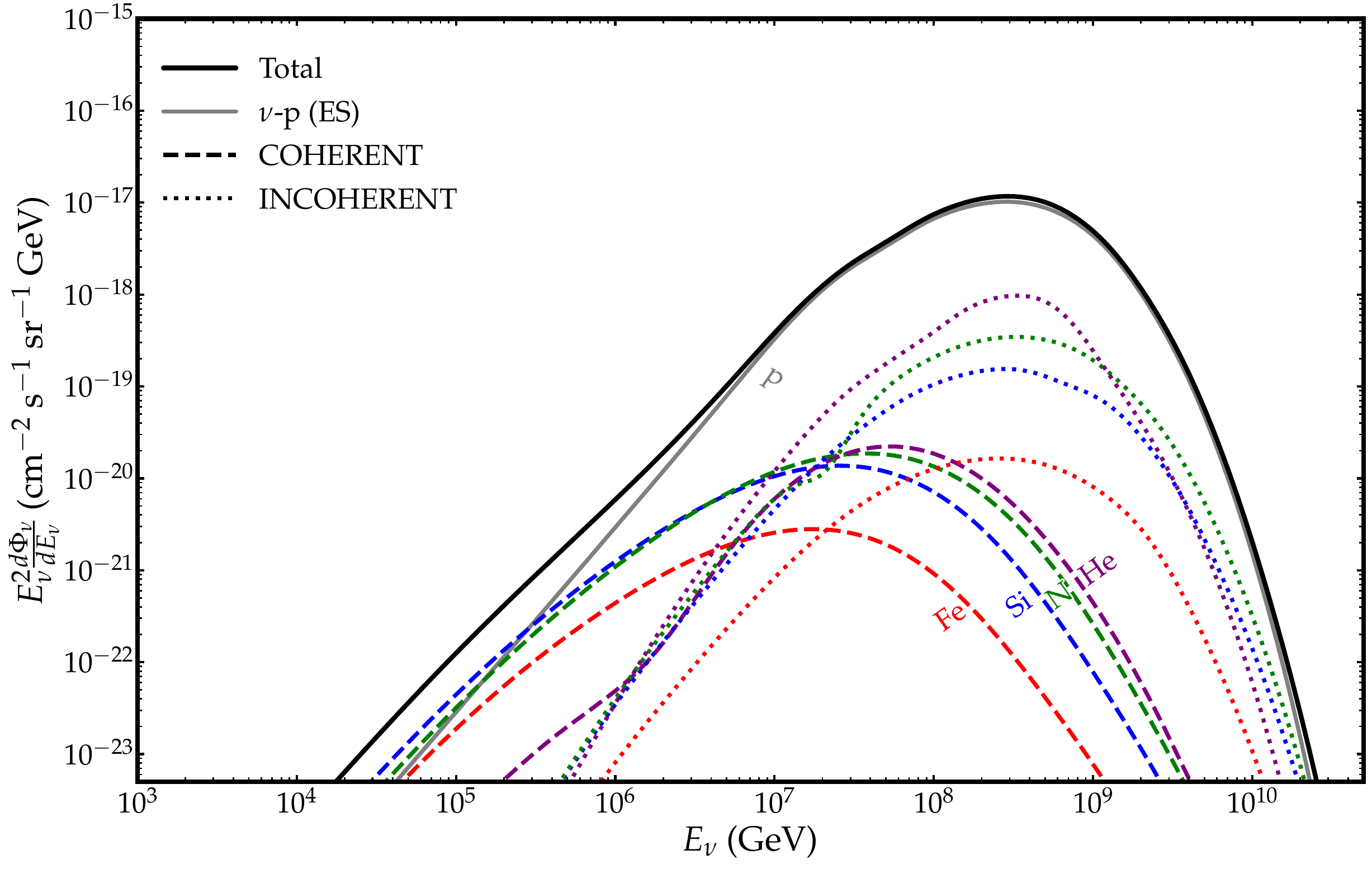}
    \includegraphics[width=0.49\textwidth]{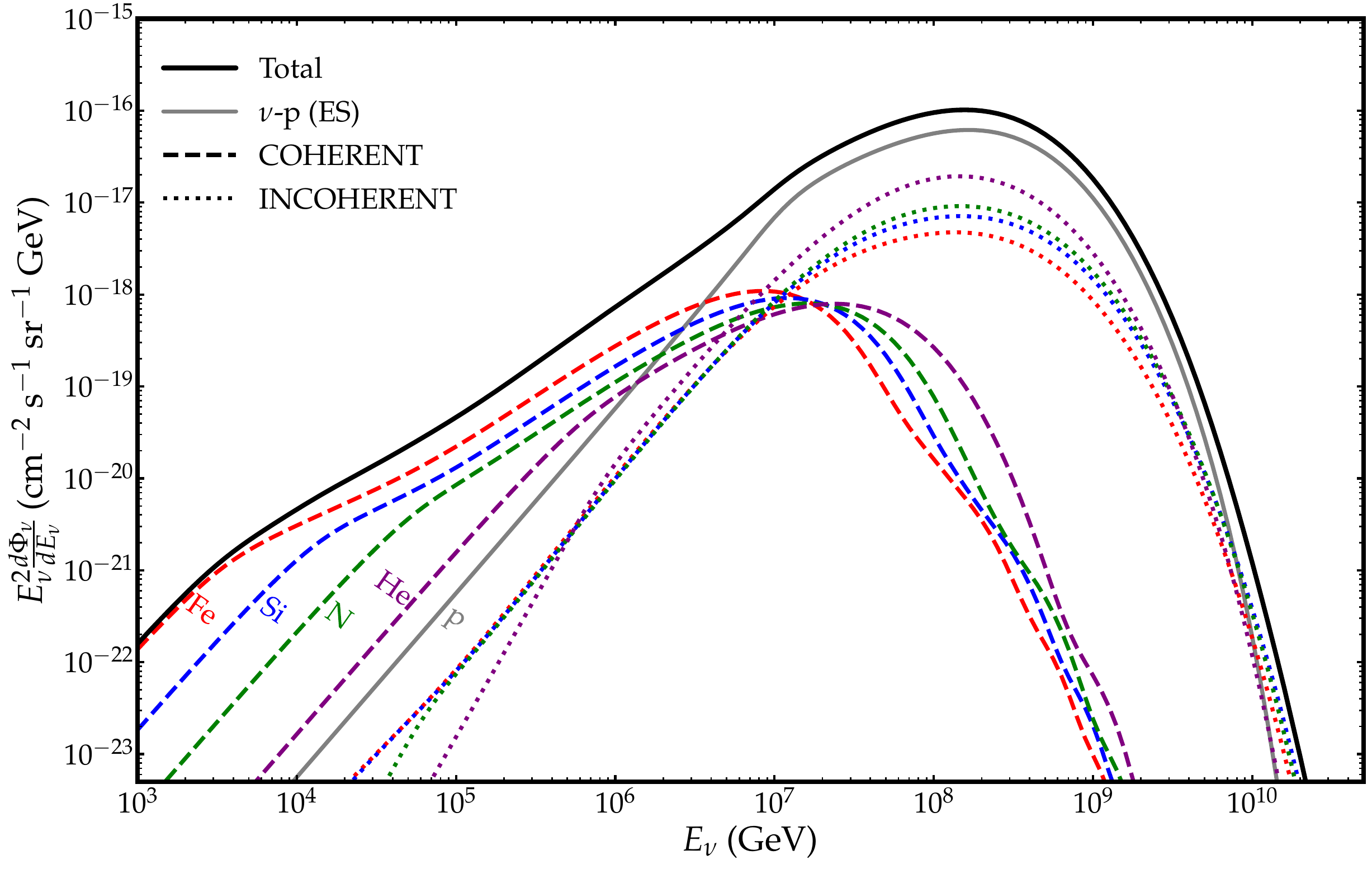}
    \caption{The boosted C$\nu$B flux as a function of neutrino energy $E_\nu$ for $\eta = 1$ and $m_1 = 0.1\,\text{eV}$. 
    Contributions from different CR nuclei are shown for comparison. 
    The upper panel uses the UHECR flux simulated with PriNCe, while the lower panel uses the Hillas model to parametrize the UHECR spectrum.
    Different colors represent various CR nuclei, with dashed and dotted lines indicating the coherent and incoherent contributions. The gray line shows the proton ES contribution, and the black line depicts the total flux. The SFR model is used for the CR source evolution.}
    \label{fig:flux}
\end{figure}

The relative contributions of different UHECR nuclei to the boosted C$\nu$B flux at Earth are shown in Fig.~\ref{fig:flux} for $m_1 = 0.1\,\text{eV}$. We assume the normal neutrino mass ordering with $\Delta m_{21}^2 = 7.42\times 10^{-5}\,\text{eV}^2$ and $\Delta m_{31}^2 = 2.51\times10^{-3}\,\text{eV}^2$~\cite{Esteban:2020cvm}; results for different $m_1$ values and the inverted ordering are provided in Appendix~D.  
In the upper panel of Fig.~\ref{fig:flux}, we show the boosted C$\nu$B flux computed using the UHECR flux obtained from the PriNCe simulations. 
For $E_\nu \lesssim 10^{6}\,\mathrm{GeV}$, the coherent scattering of heavy nuclei yields a non-negligible contribution. As $E_\nu>10^{10}\,\mathrm{GeV}$, the contribution from incoherent scattering of heavy nuclei becomes important.
The elastic scattering of protons with the C$\nu$B provides the dominant contribution to the boosted C$\nu$B flux across the entire energy range.

In the lower panel of Fig.~\ref{fig:flux}, we show the boosted C$\nu$B flux obtained from the Hillas model.
In this case, coherent scattering from heavy nuclei dominates the boosted flux at $E_\nu \lesssim 10^{7}\,\mathrm{GeV}$ due to the coherent enhancement for heavier nuclei. 
At higher energies, $E_\nu \gtrsim 10^{7}\,\mathrm{GeV}$, the contributions from proton elastic scattering and incoherent scattering of heavy nuclei become dominant.
As we can see from Fig.~\ref{fig:flux}, the predicted boosted C$\nu$B flux depends strongly on the UHECR composition. 
\begin{figure*}[t!]
    \centering
    \begin{minipage}{0.49\textwidth}
        \centering
        \includegraphics[width=\textwidth]{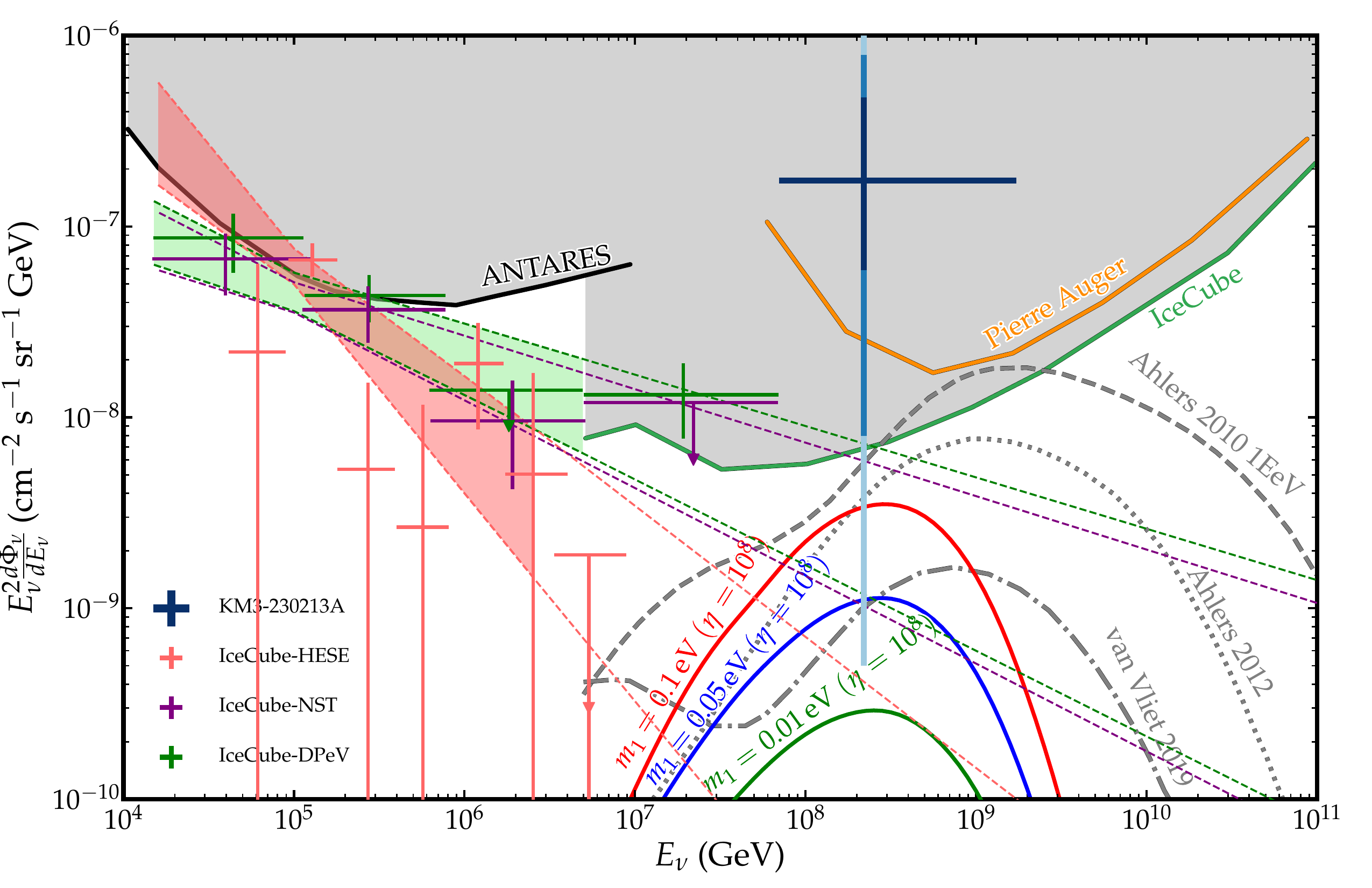}
    \end{minipage}
    \hfill
    \begin{minipage}{0.49\textwidth}
        \centering
        \includegraphics[width=\textwidth]{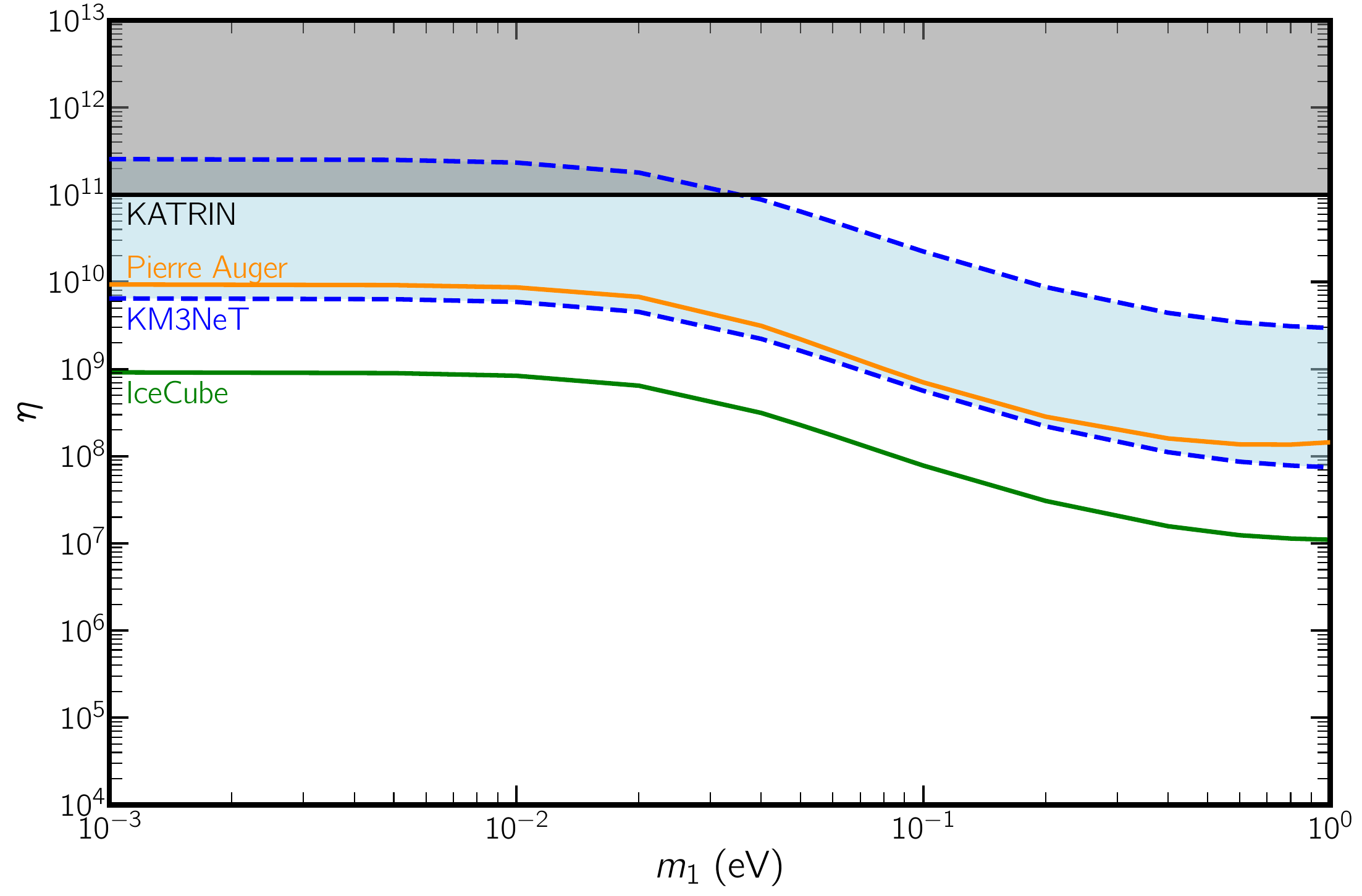}
    \end{minipage}

    \vspace{0.3cm}

    \begin{minipage}{0.49\textwidth}
        \centering
        \includegraphics[width=\textwidth]{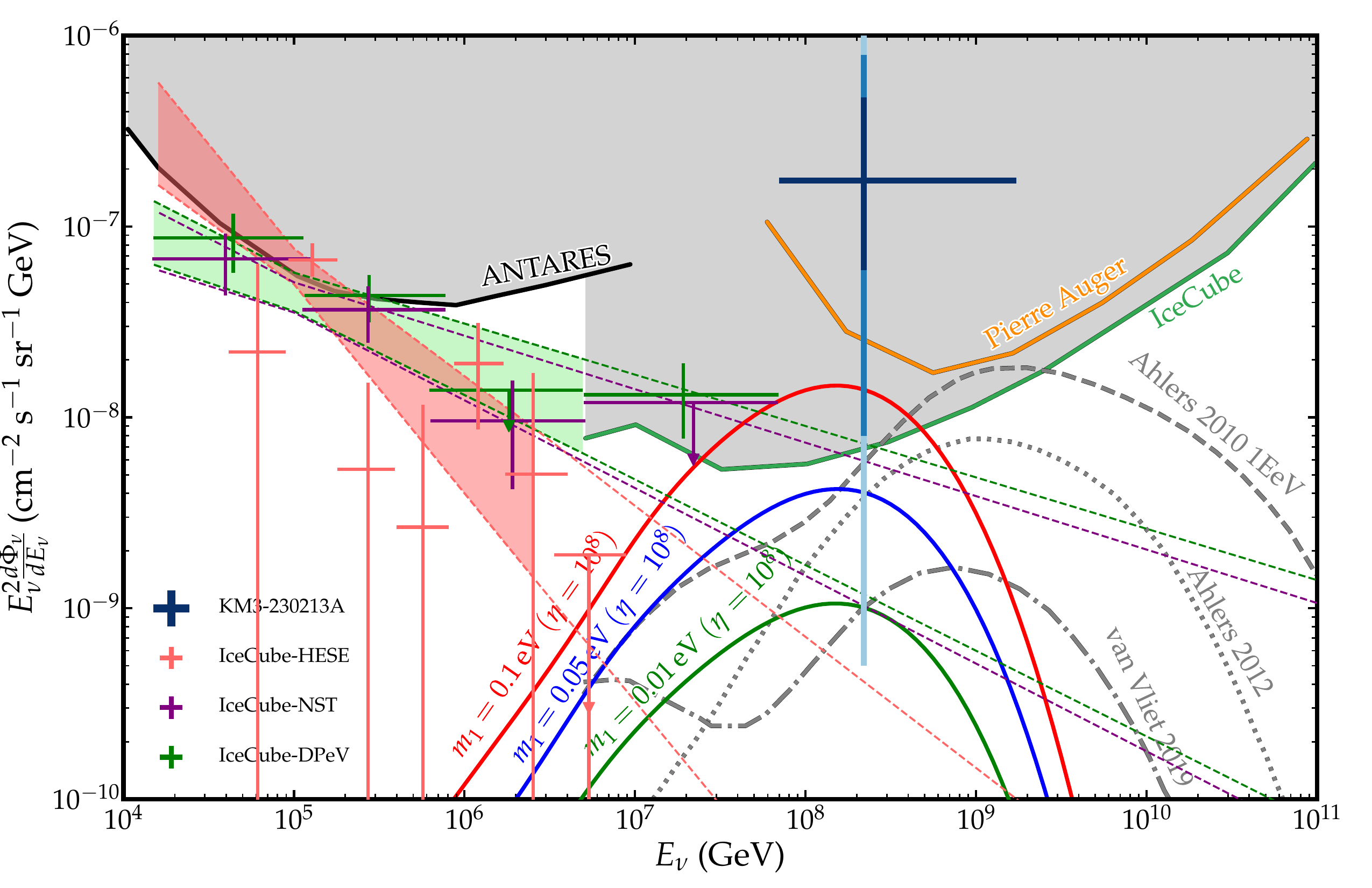}
    \end{minipage}
    \hfill
    \begin{minipage}{0.49\textwidth}
        \centering
        \includegraphics[width=\textwidth]{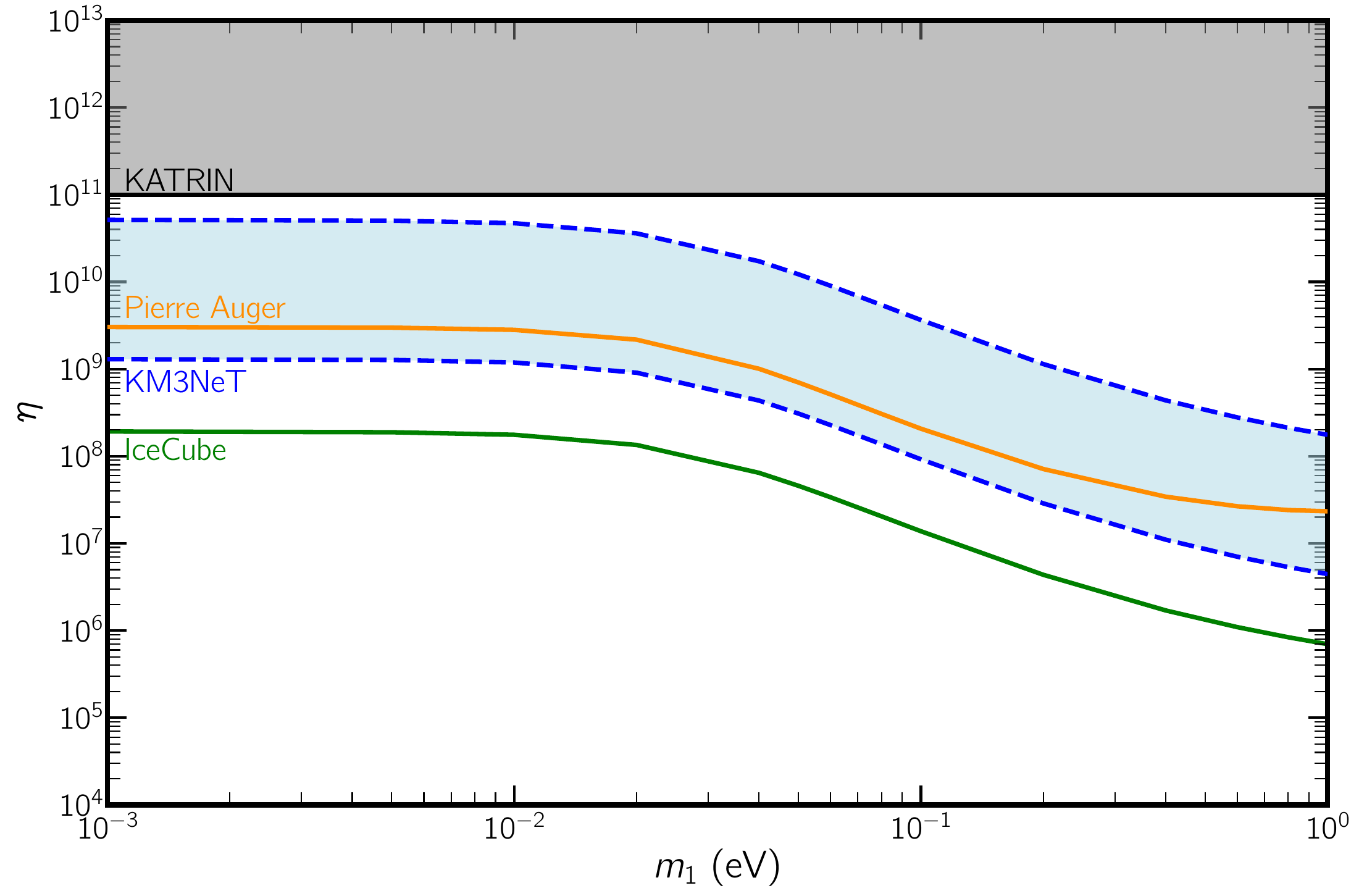}
    \end{minipage}

    \caption{
    Left panels: The all-flavor boosted C$\nu$B flux as a function of neutrino energy $E_\nu$ for different lightest neutrino masses $m_1$.  
    Red, blue, and green solid lines represent the flux for $m_1 = 0.1 \, \text{eV}$, $0.05 \, \text{eV}$, and $0.01 \, \text{eV}$, respectively.  
    Varying shades of blue correspond to the 1$\sigma$, 2$\sigma$, and 3$\sigma$ ranges for the KM3-230213A event~\cite{KM3NeT:2025npi}.  
    Also shown are limits on the diffuse cosmogenic flux from IC (90\% CL)~\cite{IceCube:2025ezc,Meier:2024flg},  
    PAO (90\% CL)~\cite{PierreAuger:2019ens}, and ANTARES (95\% CL)~\cite{ANTARES:2024ihw},  
    as well as the measured astrophysical neutrino spectrum from IC HESE~\cite{IceCube:2020wum},  
    NST~\cite{Abbasi:2021qfz}, and DPeV~\cite{IceCube:2025ary}.  
    Gray curves show representative cosmogenic neutrino flux models~\cite{Ahlers:2010fw,Ahlers:2012rz,vanVliet:2019nse}.  
    The overdensity is set to $\eta=10^8$.
    Right panels: Constraints on the C$\nu$B overdensity $\eta$ as a function of the lightest neutrino mass $m_1$.  
    The green and orange curves denote the 90\% CL constraints from IC and PAO, respectively.  
    The blue shaded region indicates the parameter space compatible with the KM3-230213A event,  
    while the gray band marks the current KATRIN exclusion at 90\% CL.  
    The upper two panels are based on the UHECR flux simulated with PriNCe, while the lower two panels are based on the UHECR flux parametrized by the Hillas model.
    All panels use the SFR model for the CR source distribution.
    }
    \label{fig:flux-SFR}
\end{figure*}
\section{Constraints on the overdensity}
In the left panels of Fig.~\ref{fig:flux-SFR}, we show the boosted C$\nu$B flux with different values of $m_1$ for an overdensity $\eta = 10^8$. We see that the boosted C$\nu$B flux decreases as the lightest neutrino mass becomes smaller. We also compare the boosted C$\nu$B flux with the current high energy neutrino data and projected sensitivity limit of various observatories. The boosted C$\nu$B flux has a peak of $\sim 200 \, \text{PeV}$, which can be probed at IC and PAO. We use IC and PAO to constrain the overdensity $\eta$ of the C$\nu$B. The expected number of events from the boosted C$\nu$B is given by
\begin{equation}
\label{eq:events}
n_{\text{events}} = 4 \pi T \int_{E_{\text{min}}}^{E_{\text{max}}} d E_\nu \frac{d \phi_\nu}{d E_\nu}(\eta, m_1) A_{\text{eff}}(E_\nu)\,,
\end{equation}
where $T$ is the data-taking time of 12.6 yr and 9.7 yr for IC~\cite{Meier:2024flg} and PAO~\cite{PierreAuger:2019ens}, respectively. $A_{\text{eff}}(E_\nu)$ is the energy-dependent effective area for all flavors extracted from Fig.~4 in Ref.~\cite{KM3NeT:2025ccp}. For IC, three neutrino events are observed at PeV energies \cite{IceCube:2025ezc,IceCube:2021rpz,2019GCN.24028....1I,IceCube:2016umi}. We conservatively take the three events as the background and obtain the 90\% CL Feldman-Cousins upper limit by requiring $n_{\text{events}}^{\text{IC}} < 1.08$~\cite{Feldman:1997qc}. For PAO, the Feldman-Cousins upper limit is set by $n_{\text{events}}^{\text{PA}}<2.39$ which takes into account the uncertainties in the exposure \cite{PierreAuger:2015ihf}.

The upper-right panel of Fig.~\ref{fig:flux-SFR} shows the constraints on the overdensity factor $\eta$ as a function of the lightest neutrino mass for the normal ordering, based on PriNCe simulations.
At $m_1 = 0.01 \, \text{eV}$, IC (PAO) imposes an upper limit on $\eta < 8.4 \times 10^8$ ($8.6 \times 10^{9}$) at the 90\% CL. For $m_1<0.01 \, \text{eV}$, the constraints become flat as the dominant contributions arise from heavier mass eigenstates with $m_2$ and $m_3$. As the lightest neutrino mass increases, the constraints on $\eta$ become more stringent.
At $m_1 = 0.1 \, \text{eV}$, the bounds from IC and PAO can reach $\eta < 7.8 \times 10^7$ and $7.0 \times 10^8$, respectively. 
As shown in the lower panels of Fig.~\ref{fig:flux-SFR}, the boosted C$\nu$B flux obtained with the Hillas model is nearly 1 order of magnitude higher than that obtained from PriNCe simulations, leading to correspondingly stronger limits on $\eta$.
The constraints for other CR source distributions 
and the inverted ordering are presented in Appendix B and Appendix D, respectively. We see that the GRB source distribution provides the strongest constraints, with approximately an order of magnitude stronger than the QSO case. For the inverted ordering, the constraints are approximately $50\%$ stronger than those for the normal ordering when the lightest neutrino mass is less than $0.01 \, \text{eV}$. 
Note that both IC and PAO set a stronger constraint than the current KATRIN upper limit of $\eta < 9.7 \times 10^{10}$ from the relic neutrino capture process~\cite{KATRIN:2022kkv}.

Recently, the KM3NeT Collaboration reported a record-breaking event KM3-230213A with an energy of $220^{+570}_{-110} \, \text{PeV}$~\cite{KM3NeT:2025npi}.Interestingly, the energy aligns with the peak of the boosted C$\nu$B flux as shown in the upper panel of Fig.~\ref{fig:flux-SFR}. This coincidence allows us to entertain the possibility of explaining the KM3NeT event by the boosted C$\nu$B from UHECR. We plot the 90\% CL allowed region that explains the KM3-230213A event in the upper-right panel of Fig.~\ref{fig:flux-SFR}. We find that explaining KM3-230213A requires an overdensity within $[5.9 \times 10^9 , 2.3 \times 10^{11}]$ for $m_1 = 0.01\,\text{eV}$, which remains below the current KATRIN bound~\cite{KATRIN:2022kkv}. Note that the allowed region is in tension with the IC bound, which is well-known and consistent with the 3.5$\sigma$ tension between IC and KM3Net for the diffuse neutrino flux as IC has a larger effective area and a longer runtime~\cite{Li:2025tqf,KM3NeT:2025ccp}. This tension can be alleviated in the presence of new physics such as sterile-to-active neutrino transitions under nonstandard matter effects~\cite{Brdar:2025azm}, or neutrino self-interactions~\cite{He:2025bex}.

With the enhanced sensitivity of upcoming telescopes such as IceCube-Gen2~\cite{Meier:2024flg}, Trinity-18~\cite{Otte:2019knb}, RNO-G~\cite{RNO-G:2023jlv}, TAMBO~\cite{Romero-Wolf:2020pzh}, POEMMA~\cite{POEMMA:2020ykm}, and GRAND~\cite{GRAND:2017pux}, detection of the boosted C$\nu$B may become possible, providing a direct evidence of the existence of relic neutrinos. However, distinguishing the boosted C$\nu$B signal from background neutrinos, such as astrophysical and cosmogenic neutrinos, remains a significant challenge. As shown in the upper panel of Fig.~\ref{fig:flux-SFR}, these background neutrinos exhibit characteristic energy spectra that differ from the boosted C$\nu$B, enabling potential discrimination with increased statistics. In addition, astrophysical and cosmogenic neutrinos are often accompanied by cascaded gamma-ray emission in the GeV to TeV range~\cite{Fang:2025nzg}, detectable via multimessenger observations with imaging air Cherenkov telescopes and air-shower gamma-ray observatories. Astrophysical neutrinos may also be associated with signatures like gravitational waves from neutron star mergers. Such multimessenger correlations could help rule out the boosted C$\nu$B hypothesis. Conversely, if no electromagnetic or gravitational-wave counterparts are observed, it could strengthen the case for identifying the neutrinos as boosted C$\nu$B.



\section{Summary}

In this paper, we study the process of the C$\nu$B boosted by UHECR by taking into account both the coherent and incoherent scattering channels. We derive the scattering cross sections for UHECR colliding with both Dirac and Majorana relic neutrinos, and show that the difference between the two cases is negligible. We find that the cross section of UHECR scattering off relic neutrinos can be coherently enhanced, which is similar to the CE$\nu$NS process observed at low-energy neutrino scattering experiments. 

We compute the boosted C$\nu$B flux produced during the propagation of UHECR from their sources to Earth.
The UHECR flux is obtained by using simulations with the UHECR propagation code PriNCe as well as a phenomenological parametrization based on the Hillas model. Compared to the PriNCe simulation, the Hillas model yields a larger fraction of heavy nuclei.
Our results show that both coherent scattering of heavy nuclei and incoherent scattering contributions to the boosted C$\nu$B flux are non-negligible.

Using current data from IC and the PAO, we set constraints on the C$\nu$B overdensity parameter $\eta$. 
Based on the results of the PriNCe simulation, we find that for $m_1 = 0.1\,\mathrm{eV}$ the bounds from IC and PAO reach $7.5 \times 10^{7}$ and $2.7 \times 10^{9}$, respectively.
Since the recent observed highest neutrino event KM3-230213A has an energy aligning with the peak of the boosted C$\nu$B flux, we also entertain the possibility of explaining the KM3-230213A event with the boosted C$\nu$B. Future observatories, such as IceCube-Gen2 and POEMMA, may enable the first detection of the boosted C$\nu$B, and potentially distinguish it from cosmogenic neutrinos through differences in the energy spectra and multimessenger observations.

\section{Acknowledgments}
We thank Karl-Heinz Kampert for useful comments and suggestions.
J.L. is supported by the National Natural Science Foundation of China under Grant No.~12275368 and the Fundamental Research Funds for the Central Universities, Sun Yat-Sen University under Grant No. 24qnpy116. A.S. acknowledges funding from the German Bundesministerium für Bildung und Forschung (F\"{o}rderkennzeichen 05A23PX3).

\section*{DATA AVAILABILITY}
 The data that support the findings of this article are openly available\footnote{\url{https://github.com/jiajie-z999/UHECR-boosted-relic-neutrinos.git}}.


\clearpage
\appendix
\onecolumngrid 

\section{CROSS SECTIONS}

{\bf Kinematics.} To analyze the scattering of a CR nucleus $\mathcal{N}_i$ off a relic neutrino at rest, we define the four–momenta as follows: 
$P_{\nu} = (E,\vec p) \simeq (m_{\nu}, 0)$ for the incoming neutrino,
and $P_{\nu}' = \{E_{\nu}, \vec{p}_{\nu}\}$ for the outgoing neutrino; $K_{\mathcal{N}} = \{E_{\mathcal{N}}, \vec{p}_{\mathcal{N}}\}$ for the initial nucleus and $K_{\mathcal{N}}' = \{E_{\mathcal{N}}', \vec{p}_{\mathcal{N}}'\}$ for the final nucleus. Given the energy scales involved, we apply the approximation $E_{\mathcal{N}_i} \gg m_{\mathcal{N}_i} \gg m_\nu$, which holds for UHECR interactions. Using energy–momentum conservation, the Mandelstam variables are approximated as given by
\begin{equation}
\label{eq:Mandelstam}
s \approx 2 m_\nu E_{\mathcal{N}} + m_{\mathcal{N}}^2, \quad t = -q^2 \approx -2 m_\nu E_\nu, \quad u \approx m_{\mathcal{N}}^2 - 2 m_\nu (E_{\mathcal{N}} - E_\nu)\,,
\end{equation}
The maximum energy of the scattered neutrino, $E_\nu^{\max}$, is given by
\begin{equation}
\label{eq:E-max}
E_\nu^{\max}(E_{\mathcal{N}_i}) = \frac{E_{\mathcal{N}_i}^2}{E_{\mathcal{N}_i} + m_{\mathcal{N}_i}^2 / (2 m_\nu)} \approx \begin{cases} E_{\mathcal{N}_i} & \text{if } E_{\mathcal{N}_i} \gg \frac{m_{\mathcal{N}_i}^2}{2 m_\nu}, \\ \frac{2 m_\nu E_{\mathcal{N}_i}^2}{m_{\mathcal{N}_i}^2} & \text{if } E_{\mathcal{N}_i} \ll \frac{m_{\mathcal{N}_i}^2}{2 m_\nu}\,,
\end{cases}
\end{equation}

{\bf Coherent neutrino–nucleus scattering cross sections.} 
Before deriving the scattering cross sections, we first clarify the helicity and chirality properties of the C$\nu$B.
Because the weak interaction couples only to left-chiral fermions, neutrinos are produced in left-chiral states, while antineutrinos are produced in right-chiral states.
During subsequent free streaming, helicity is conserved, whereas chirality is not for massive neutrinos, since the neutrino mass will mix the left- and right-chiral components under free evolution~\cite{Long:2014zva}.
For clarity, we use helicity eigenstates to describe the neutrino and antineutrino spinors, i.e.,
\begin{equation}
u(\vec p,h)=
\begin{pmatrix}
\sqrt{E-h|\vec p|}\,\xi_h\\[2pt]
\sqrt{E+h|\vec p|}\,\xi_h
\end{pmatrix},\qquad
v(\vec p,h)=
\begin{pmatrix}
h\sqrt{E+h|\vec p|}\,\xi_{-h}\\[2pt]
-h\,\sqrt{E-h|\vec p|}\,\xi_{-h}
\end{pmatrix}\,,
\end{equation}
where $(\boldsymbol{\sigma}\!\cdot\!\hat{\vec p})\,\xi_h = h\,\xi_h$, $\boldsymbol{\sigma}$ denotes the Pauli matrices, $\hat{\vec p} = \vec p / |\vec p|$, and $\xi_h$ is a two-component Weyl spinor with definite helicity $h$. 
The spinors are normalized as $\xi_h^\dagger \xi_{h'} = \delta_{hh'}$, and $h = \pm 1$ corresponds to right– and left–helical states, respectively.
For a neutrino in a helicity eigenstate, the left- and right-chiral components are given by
$u_{hL} = P_L u$ and $u_{hR} = P_R u$, and their relative weights are
\begin{equation}
f_{hL}=\frac{E-h|\vec p|}{2E}=\frac{1-h\beta}{2}\,,\qquad 
f_{hR}=\frac{1+h\beta}{2}\,,
\end{equation}
with $\beta = |\vec p|/E$. 
In the ultrarelativistic limit ($\beta \simeq 1$), a left–helical neutrino ($h = -1$) is predominantly left chiral, whereas in the nonrelativistic limit ($\beta \simeq 0$) it becomes an approximately equal admixture of left– and right–chiral components. 
The C$\nu$B neutrinos today are in the nonrelativistic regime, and since only the left–chiral component participates in weak interactions, the scattering amplitude is suppressed by the corresponding weight factor $f_{hL}$.

We first compute the coherent elastic neutrino-nucleus scattering cross section for a C$\nu$B composed of Dirac neutrinos. 
The scattering cross sections for left–helical neutrinos and right–helical antineutrinos are evaluated separately. 
We assume a lepton–symmetric cosmology, such that neutrinos and antineutrinos have equal number densities~\cite{Grohs:2020xxd,Long:2014zva}.
For coherent scattering, the axial–vector nuclear responses can be neglected, and we treat the nucleus as a spin–0 scalar target~\cite{Lindner:2016wff}. 
Assuming an incoming neutrino with helicity $h$ and an outgoing neutrino with helicity $h'$, the tree-level amplitudes for neutrino and antineutrino scattering in the Dirac case are
\begin{equation}
i\mathcal{M}_{D}^{\nu}
= i \frac{\sqrt{2}}{2}\,G_F\,Q_W\,F(q^2)\,(K_{\mathcal N_i}+K'_{\mathcal N_i})_\mu\;
\bar u(P',h')\,\gamma^\mu(1-\gamma^5)\,u(P,-)\,, 
\end{equation}
\begin{equation}
i\mathcal{M}_{D}^{\bar\nu}
= i \frac{\sqrt{2}}{2}\,G_F\,Q_W\,F(q^2)\,(K_{\mathcal N_i}+K'_{\mathcal N_i})_\mu\;
\bar v(P,+)\,\gamma^\mu(1-\gamma^5)\,v(P',h')\,, 
\end{equation}
The nuclear weak charge is given by $Q_{W,i} = Z_i g_V^p + N_i g_V^n$, with 
$g_V^p = \tfrac{1}{2} - 2 \sin^2 \theta_W$, 
$g_V^n = -\tfrac{1}{2}$, 
and $\sin^2 \theta_W = 0.231$~\cite{ParticleDataGroup:2024cfk}. 
The polarized spin–density matrices take the form 
$u(P,-)\,\overline{u}(P,-)=\tfrac{1}{2}\,(\slashed P + m_\nu)\,\bigl(1 + \gamma^5 \slashed S\bigr)$ and 
$v(P,+)\,\overline{v}(P,+)=\tfrac{1}{2}\,(\slashed P - m_\nu)\,\bigl(1 + \gamma^5 \slashed S\bigr)$, 
where the polarization four–vector is 
$S^\mu = h\left(\tfrac{|\vec p|}{m_\nu},\, \tfrac{E}{m_\nu}\,\hat{\vec p}\right)$, 
satisfying $S\!\cdot\!P = 0$ and $S^2 = -1$. 
In the nonrelativistic regime $\beta \ll 1$, these expressions reduce to the simplified forms given below,
\begin{equation}
u(P,-)\,\overline{u}(P,-)\;\simeq\;\tfrac{m_\nu}{2}\,(\gamma^0+1)\,\bigl(1+\gamma^5\,\hat{\vec p}\cdot\vec\gamma\bigr)\,,\qquad
v(P,+)\,\overline{v}(P,+)\;\simeq\;\tfrac{m_\nu}{2}\,(\gamma^0-1)\,\bigl(1-\gamma^5\,\hat{\vec p}\cdot\vec\gamma\bigr)\,.
\end{equation}
For an isotropic relic neutrino background, terms linear in $\hat{\vec p}$ vanish upon angular averaging. 
As a result, the spin density matrices reduce to the compact forms 
$u(P,-)\overline{u}(P,-)\simeq \tfrac{m_\nu}{2}(\gamma^0+1)$ and 
$v(P,+)\overline{v}(P,+)\simeq \tfrac{m_\nu}{2}(\gamma^0-1)$. 
For the unobserved final-state helicity, we employ the completeness relations 
$\sum_{h'=\pm}u(P',h')\,\overline{u}(P',h')=\slashed P'+m_\nu$ and 
$\sum_{h'=\pm}v(P',h')\,\overline{v}(P',h')=\slashed P'-m_\nu$.
After taking the Dirac traces, all terms that distinguish the $\nu$ and $\bar{\nu}$ leptonic tensors vanish because they contain an odd number of $\gamma$ matrices, and the two tensors become identical,
\begin{equation}
\begin{split}
\label{eq:The leptonic tensors-Dirac}
L^{\mu\nu}_D
&=\mathrm{Tr}\!\left[u(P,-)\,\overline{u}(P,-)\,
\gamma^\mu(1-\gamma^5)(\slashed P'\!\pm m_\nu)\,
\gamma^\nu(1-\gamma^5)\right] \\
&= 4m_\nu\left(\delta^\mu{}_0\,P'^{\,\nu}
+\delta^\nu{}_0\,P'^{\,\mu}-g^{\mu\nu}P'^0\right)
+4i\,m_\nu\,\epsilon^{\mu\nu\alpha0}\,g_{\alpha0}\,.
\end{split}
\end{equation}
The antisymmetric $\epsilon^{\mu\nu\alpha\beta}$ term vanishes upon contraction with the symmetric nuclear current, so that only the symmetric part contributes. 
As a result, the squared amplitudes are identical for neutrino and antineutrino scattering.
\begin{equation}
|\mathcal{M}_D^{\nu}|^2=|\mathcal{M}_D^{\bar\nu}|^2
=16\,G_F^2\,Q_W^2\,F^2(q^2)\,m_\nu^2 E_{\mathcal N_i}^2
\left(1-\frac{E_\nu}{E_{\mathcal N_i}}-\frac{m_{\mathcal N_i}^2 E_\nu}{2 m_{\nu}E_{\mathcal N_i}^2}\right)\,.
\label{eq:MDirac}
\end{equation}

We now consider the coherent elastic neutrino-nucleus scattering cross section in the Majorana case. 
For a Majorana field satisfying $\nu = \nu^c$, the vector current vanishes, $\bar{\nu}\gamma^\mu\nu = 0$, while the axial current remains nonzero, $\bar{\nu}\gamma^\mu\gamma^5\nu \neq 0$. 
As a result, the leptonic vertex is purely axial and acquires an additional factor of 2 at the amplitude level compared to the Dirac case,
\begin{equation}
i\mathcal M_{\rm M}^\nu
= i\sqrt{2}\,G_F\,Q_W\,F(q^2)\,(K_{\mathcal N_i}+K'_{\mathcal N_i})_\mu\;
\overline{u}(P',h')\,\gamma^\mu\gamma^5\,u(P,h)\,.
\end{equation}
Since both helicity states are present in the present-day C$\nu$B, we average over the initial helicities,
$\tfrac{1}{2}\sum_h u(P,h)\,\overline{u}(P,h)=\tfrac{1}{2}(\slashed P+m_\nu)$,
which in the extreme nonrelativistic limit reduces to
$\tfrac{1}{2}(\slashed P+m_\nu)\simeq \tfrac{m_\nu}{2}(\gamma^0+1)$.
For the unobserved final-state helicity, we use the completeness relation,
$\sum_{h'} u(P',h')\,\overline{u}(P',h')=\slashed P'+m_\nu$.
The corresponding leptonic tensor is
\begin{equation}
L^{\mu\nu}_{\rm M}
= 2 m_\nu\left(\delta^\mu{}_0\,P'^{\,\nu}
+\delta^\nu{}_0\,P'^{\,\mu}
-g^{\mu\nu}P'^0
+m_\nu\,g^{\mu\nu}\right)\,, 
\end{equation}
and the squared amplitude in the Majorana case reads
\begin{equation}
\begin{split}
|\mathcal M_{\rm M}^\nu|^2
&=2\,G_F^2\,Q_W^2\,F^2(q^2)\,(K_{\mathcal N_i}+K'_{\mathcal N_i})_\mu\,
(K_{\mathcal N_i}+K'_{\mathcal N_i})_\nu\,L^{\mu\nu}_{\rm M} \\
&=16\,G_F^2\,Q_W^2\,F^2(q^2)\,m_\nu^2\,E_{\mathcal N_i}^2
\left(1-\frac{E_\nu}{E_{\mathcal N_i}}
-\frac{m_{\mathcal N_i}^2 E_\nu}{2 m_{\nu}E_{\mathcal N_i}^2}+\mathcal O\!\left(\frac{m_{\mathcal N_i}^{\,2}}{E_{\mathcal N_i}^2}\right)\right)\,.
\end{split}
\label{eq:MMajor}
\end{equation}
Note that we have taken the differential scattering cross section of Dirac neutrinos as the sum of neutrino and antineutrino contributions,
$d\sigma^{\nu \mathcal{N}_i}/dE_\nu \equiv
(d\sigma^{\nu \mathcal{N}_i}/dE_\nu + d\sigma^{\bar{\nu} \mathcal{N}_i}/dE_\nu)$.
For Majorana neutrinos, both helicity states are present in the C$\nu$B. Since the number density of Majorana neutrinos equals the sum of the left-helical neutrino and right-helical antineutrino densities in the Dirac case~\cite{Long:2014zva}, the scattering cross section for the Majorana case is taken to be twice that of an averaged helical neutrino contribution.

Comparing Eqs~(\ref{eq:MDirac}) and~(\ref{eq:MMajor}), we can see that the cross sections between Dirac and Majorana neutrinos only differ by a term proportional to 
$\mathcal{O}(m_{\mathcal{N}_i}^{\,2}/E_{\mathcal{N}_i}^{2})$.
It is well-known that, in the rest frame of nucleus,
the difference between the cross sections of Dirac and Majorana neutrinos is generically suppressed by $\mathcal{O}(m_\nu/E_\nu)^2$.
Our result is consistent with this expectation. Indeed, after performing a Lorentz transformation to the nuclear rest frame,
where $E_\nu = (E_{\mathcal{N}_i}/m_{\mathcal{N}_i})\, m_\nu$, the correction term $\mathcal{O}(m_{\mathcal{N}_i}^{\,2}/E_{\mathcal{N}_i}^{2})$
is seen to correspond directly to $\mathcal{O}(m_\nu/E_\nu)^2$.
Since $m_{\mathcal{N}_i}\ll E_{\mathcal{N}_i}$, we see that the difference between the cross sections of Dirac and Majorana neutrinos are negligible.

In the following analysis, we consider Dirac neutrinos and calculate the corresponding differential scattering cross section as
\begin{equation}
\label{eq:coherent-Dirac}
\frac{d \sigma_{\text{coh}}^{\nu \mathcal N_i}}{d E_{\nu}} 
= \frac{G_{F}^2 m_{\nu}}{\pi}\, F^2(q^2)\, Q_{W_i}^2 
\left( 1 - \frac{E_{\nu}}{E_{\mathcal N_i}} - \frac{m_{\mathcal N_i}^2 E_{\nu}}{2 m_{\nu} E_{\mathcal N_i}^2} \right)\,.
\end{equation}
It should be emphasized that the coherent scattering cross section derived here cannot be obtained by a simple Lorentz transformation of the CE$\nu$NS cross sections derived in the rest frame of nucleus, such 
as those presented in Ref.~\cite{Lindner:2016wff}.
This is because chirality is Lorentz invariant. In Ref.~\cite{Lindner:2016wff}, the incoming neutrinos are ultra-relativistic and therefore, effectively left-chiral, whereas today’s C$\nu$B neutrinos are nonrelativistic helicity eigenstates containing both left- and right-chiral components.
Consequently, applying a Lorentz boost to CE$\nu$NS cross sections derived in the nuclear rest frame leads to an overestimation of the coherent scattering cross section by approximately a factor of 2.

Also note that this behavior contrasts with the case of C$\nu$B capture, where the capture rate for Majorana neutrinos is twice that for Dirac neutrinos~\cite{Long:2014zva,Roulet:2018fyh}. The origin of this factor of 2 enhancement is that, in the capture process, only neutrinos can be captured, while antineutrinos are kinematically forbidden.
In our scenario, both neutrinos and antineutrinos in the C$\nu$B can be boosted by UHECRs and contribute symmetrically to the scattering process. As a result, no such factor of 2 difference arises between the Dirac and Majorana cases.

{\bf Elastic neutrino–nucleon scattering cross section.}
Elastic neutrino-nucleon scattering has been extensively studied in the literature~\cite{Giunti:2007ry,Formaggio:2012cpf,DeMarchi:2024zer}, focusing on the regime where the incoming neutrino is ultrarelativistic.
The corresponding scattering cross sections derived in these works~\cite{Giunti:2007ry,Formaggio:2012cpf,DeMarchi:2024zer} cannot be directly applied to UHECR--C$\nu$B scattering by a Lorentz transformation.
As discussed above, the difference between Dirac and Majorana neutrinos is negligible, and we therefore work in the Dirac framework. The leptonic tensor takes the form given in Eq.~\eqref{eq:The leptonic tensors-Dirac}.
The corresponding hadronic tensor is given by
\begin{equation}
H^{\mu\nu}=\frac{1}{2}\,\mathrm{Tr}\!\left[(\slashed K+m_N)\,\Gamma_Z^\mu(q)\,(\slashed K'+m_N)\,\overline{\Gamma}_Z^{\,\nu}(q)\right]\,,
\end{equation}
where the neutral-current nucleon vertex is given by
\begin{equation}
\Gamma_Z^\mu(q)=F_1^{ZN}(q^2)\,\gamma^\mu+\frac{i}{2m_N}F_2^{ZN}(q^2)\,\sigma^{\mu\nu}q_\nu - G_A^{ZN}(q^2)\,\gamma^\mu\gamma^5\,,
\end{equation}
where $F_1^{ZN}$ and $F_2^{ZN}$ are the Dirac and Pauli electromagnetic form factors for the weak neutral currents, and they are defined as $F_{i}^{ZN} = \pm \frac{1}{2} (F_{i}^{p} - F_{i}^{n}) - 2 \sin^2 \theta_W F_{i}^{N}$, with the $+(-)$ sign for $N = p(n)$. The axial form factor is $G_A^{ZN} = \pm (1/2) G_A$, with the same sign convention. The electric and magnetic form factors are defined as~\cite{Giunti:2007ry}:
\begin{equation}
G_E^N(q^2) = F_1^N(q^2) - \frac{q^2}{4 m_N^2} F_2^N(q^2), \quad G_M^N(q^2) = F_1^N(q^2) + F_2^N(q^2)\,,
\end{equation}
with $q^2$ dependence given by $G_{E,M}^N(q^2) = G_{E,M}^N(0) (1 + q^2 / \Lambda_{E,M}^2)^{-2}$, where $\Lambda_{E,M} \simeq 0.8 \, \text{GeV}$, and $G_A(q^2) = G_A(0) (1 + q^2 / m_A^2)^{-2}$, with $G_A(0) \simeq 1.245$ and $m_A \simeq 1.17 \, \text{GeV}$~\cite{Gao:2021sml,Alexandrou:2023qbg}. At zero momentum transfer ($q^2 = 0$), these form factors become $G_E^p(0) = 1$, $G_E^n(0) = 0$, $G_M^p(0) = \mu_p/\mu_N$, and $G_M^n(0) = \mu_n/\mu_N$, where $\mu_p/\mu_N = 2.79$ and $\mu_n/\mu_N = -1.91$~\cite{ParticleDataGroup:2024cfk} are the magnetic moments of the proton and neutron, respectively.

The trace is evaluated using {\tt FeynCalc}~\cite{Shtabovenko:2020gxv} and the
kinematic relations $E_\nu = q^2/(2m_\nu)$ and
$E_N = (s-2m_N^2)/(2m_N)$ in the laboratory frame.
The resulting expression has the same functional dependence as those obtained in
Refs.~\cite{Giunti:2007ry,Formaggio:2012cpf}, but differs by an overall factor of 2.
The elastic differential cross section in the laboratory frame is
\begin{equation}
\label{ES-cross section}
\frac{d \sigma_{\text{ES}}^{\nu N}}{d E_\nu} = \frac{G_F^2\, m_\nu m_N^4}{2\pi\,(s - m_N^2)^2}\,
\left[ A(q^2) \pm B(q^2)\,\frac{s-u}{m_N^2} + C(q^2)\,\frac{(s-u)^2}{m_N^4} \right]\,,
\end{equation}
where the $\pm$ sign is for $\nu/\bar\nu$. The coefficient functions in terms of the neutral–current form factors are
\begin{equation}
A(q^2)=\frac{q^2}{m_N^2}\!\left\{\!\left(1+\frac{q^2}{4m_N^2}\right)\!\bigl(G_A^{Z}\bigr)^2
-\left(1-\frac{q^2}{4m_N^2}\right)\!\left[\bigl(F_1^{Z}\bigr)^2-\frac{q^2}{4m_N^2}\bigl(F_2^{Z}\bigr)^2\right]
+\frac{q^2}{m_N^2}\,F_1^{Z}F_2^{Z}\right\}\!,
\end{equation}
\begin{equation}
B(q^2)=\frac{q^2}{m_N^2}\,G_A^{Z}\,\bigl(F_1^{Z}+F_2^{Z}\bigr)\,,
\end{equation}
\begin{equation}
C(q^2)=\frac{1}{4}\left[\bigl(G_A^{Z}\bigr)^2+\bigl(F_1^{Z}\bigr)^2+\frac{q^2}{4m_N^2}\bigl(F_2^{Z}\bigr)^2\right]\,.
\end{equation}

{\bf Incoherent neutrino-nucleus scattering cross section.} 
The squared incoherent amplitude for neutrino-nucleus scattering can be written as a linear sum of the squared neutrino-nucleon amplitudes multiplied by the factor $1 - F^2(q^2)$ [see Eq.~(18) in Ref.~\cite{Bednyakov:2018mjd}], i.e.,
\begin{equation}
|\mathcal{A}|^2_{\text{incoh}} = \left( Z_i |\mathcal{A}_k^p|^2 + N_i |\mathcal{A}_j^n|^2 \right) (1 - F^2(q^2))\,,
\label{eq:incoherent_amplitude}
\end{equation}
where $\mathcal{A}_k^p$ ($\mathcal{A}_j^n$) is the amplitude for neutrinos scattering off individual proton (neutron), and $F(q^2)$ is the nuclear form factor for single nucleons. Therefore, the incoherent differential cross section for neutrino-nucleus scattering is
\begin{equation}
\label{eq:incoherent2}
\frac{d \sigma^{\nu \mathcal{N}_i}_\text{incoh}}{d E_\nu} = \left[ Z_i \frac{d \sigma_{\text{ES}}^{\nu p}}{d E_\nu} + N_i \frac{d \sigma_{\text{ES}}^{\nu n}}{d E_\nu} \right] (1 - F^2(q^2))\,,
\end{equation}
where $d \sigma_{\text{ES}}^{\nu p}/d E_\nu$ and $d \sigma_\text{ES}^{\nu n}/d E_\nu$ are the differential cross sections for neutrino-proton and neutrino-neutron scattering, as given in Eq.~\eqref{ES-cross section}. 
For a small momentum transfer $q \ll \Lambda_{E,M} (\text{or }m_A)$, 
the $q$ dependence of nucleon form factors can be ignored, allowing the approximation $G_A(q^2) = G_A(0)$ and $G_{E,M}^N(q^2) = G_{E,M}^N(0)$. The coefficients in Eq.~\eqref{ES-cross section} can be simplified to
\begin{equation}
A_N(0) = \frac{q^2}{m_N^2} \left[ G_{A}^{ZN}(0)^2 - F_1^{ZN}(0)^2 \right], \quad C_N(0) = \frac{1}{4} \left[ G_{A}^{ZN}(0)^2 + F_1^{ZN}(0)^2 \right]\,,
\label{simplifying_AC}
\end{equation}
where $G_A^{ZN}(0) = \pm \frac{1}{2} G_A(0) = g_A^N$ and $F_1^{ZN}(0) = \pm \frac{1}{2} (F_1^p(0) - F_1^n(0)) - 2 \sin^2 \theta_W F_1^N(0) = g_V^N$. The coefficient $B_N = 0$ due to cancellation between neutrino and antineutrino contributions. Using the Mandelstam variables from Eq.~\eqref{eq:Mandelstam} and the simplified
coefficients $A_N(0)$ and $C_N(0)$ from Eq.~\eqref{simplifying_AC},
the elastic scattering cross section in Eq.~\eqref{eq:incoherent2}.
After summing the contributions from neutrinos and antineutrinos,
the incoherent scattering cross section is given by
\begin{equation}
\label{eq:incoherent_derived}
\frac{d\sigma^{\nu \mathcal{N}_i}_\text{incoh}}{d E_\nu} = \frac{ G_F^2 m_\nu}{\pi} \left[ (Z_i (g_V^p)^2 + N_i (g_V^n)^2) \left( 1 - \frac{E_\nu}{E_N} - \frac{m_N^2 E_\nu}{2 m_\nu E_N^2} \right) + (Z_i (g_A^p)^2 + N_i (g_A^n)^2) \left( 1 - \frac{E_\nu}{E_N} + \frac{m_N^2 E_\nu}{2 m_\nu E_N^2} \right) \right] (1 - F^2(q^2))\,.
\end{equation}
In the present analysis, the relic neutrinos are nonrelativistic, whereas Ref.~\cite{Bednyakov:2018mjd} considered the ultrarelativistic limit.
Our result corresponds to one half of the expression for unpolarized nuclei
given in Eq.~(46) of Ref.~\cite{Bednyakov:2018mjd}, which reads
\begin{equation}
\label{eq:Bednyakov}
\frac{d\sigma^{\nu \mathcal{N}_i}_\text{incoh}}{d E_\nu} = g_{\text{inc}} \frac{2 G_F^2 m_\nu}{\pi} \sum_{N=p,n} A^N \left[ (g_L^N)^2 + (g_R^N)^2 (1 - y)^2 - \frac{2 g_L^N g_R^N m_N y}{s - m_N^2} \right] (1 - F^2(q^2))\,,
\end{equation}
where $A^p = Z_i$ and $A^n = N_i$ are the numbers of protons and neutrons in the nucleus $\mathcal{N}_i$, respectively. The variable $y = \frac{(P_{\nu} \cdot q)}{(P_{\nu} \cdot K_{\mathcal{N}})} = \frac{s - m_N^2}{s} \frac{q^2}{q_{\text{max}}^2} = \frac{E_\nu}{E_N}$. The incoherent correction factors are given by $g_{\text{inc}} = C^N_1 C^N_2C^N_3$~\cite{Bednyakov:2018mjd,Bednyakov:2021lul}. Here, $C^N_1$ is the correction factor arises from the transition from the nucleus (a multiparticle state) to nucleons (single-particle states) in the scattering amplitude calculation. The expression of $C^N_1$ is given in Eq.~(82) of Ref.~\cite{Bednyakov:2021lul}, with a value of about unity~\cite{Bednyakov:2021lul}. $C^N_2$ is a kinematic correction factor related to the nuclear excitation energy $\Delta \epsilon$, with its expression provided in Eq.~(96) of Ref.~\cite{Bednyakov:2021lul}. For iron nuclei, since the excitation energies are $\sim \mathcal{O}(\text{MeV})$ and significantly smaller than the nuclear mass $\sim \mathcal{O}(50 \, \text{GeV})$, the value of $C^N_2$ is approximately unity~\cite{Bednyakov:2023pwi,Bednyakov:2021lul}. 
The correction factor $C^N_3$ takes into account the nonzero three momentum for the
nucleon in the initial state, and the value can be taken to be unity in the leading approximation~\cite{Bednyakov:2021lul}. 
Consequently, we adopt $g_{\text{inc}}\approx 1$. For antineutrinos, the couplings are flipped ($g_L^N \to g_R^N$, $g_R^N \to g_L^N$), yielding
\begin{equation}
\frac{d\sigma^{\bar{\nu} \mathcal{N}_i}_\text{incoh}}{d E_\nu} = g_{\text{inc}} \frac{2 G_F^2 m_\nu}{\pi} \sum_{N=p,n} A^N \left[ (g_R^N)^2 + (g_L^N)^2 (1 - y)^2 - \frac{2 g_L^N g_R^N m_N y}{s - m_N^2} \right] (1 - F^2(q^2))\,.
\end{equation}
Summing over the contributions of both neutrino and antineutrino, the differential cross section becomes
\begin{equation}
\label{eq:incoherent-LR}
\frac{d\sigma^{\nu \mathcal{N}_i}_\text{incoh}}{d E_\nu} \equiv \left( \frac{d\sigma^{\nu \mathcal{N}_i}_\text{incoh}}{d E_\nu} + \frac{d\sigma^{\bar{\nu} \mathcal{N}_i}_\text{incoh}}{d E_\nu} \right) = \frac{2 G_F^2 m_\nu}{\pi} \sum_{N=p,n} A^N \left[ ((g_L^N)^2 + (g_R^N)^2) \left( 1 + (1 - y)^2 \right) - \frac{4 g_L^N g_R^N m_N y}{s - m_N^2} \right] (1 - F^2(q^2))\,.
\end{equation}
Using the vector (axial-vector) couplings $g_V^N = g_L^N + g_R^N$ ($g_A^N = g_L^N - g_R^N$) and $y=\tfrac{E_\nu}{E_N}$, it is straightforward to show that Eq.~\eqref{eq:incoherent-LR} is the same as Eq.~\eqref{eq:incoherent_derived}. 
Neglecting the $q$ dependence of the nucleon form factors, our incoherent cross section reduces to one half of the unpolarized-nucleus expression given in Eq.~(46) of Ref.~\cite{Bednyakov:2018mjd}.
Note that Ref.~\cite{Bednyakov:2018mjd} considered accelerator neutrinos with energies $\sim\mathcal{O}(10)$ MeV scattering off nuclear targets at rest, where the momentum transfers are very small, i.e.\ $q^2 / \Lambda_{E,M}^2(m_A^2) \approx 0$, allowing the approximation $G_A(q^2) = G_A(0)$ and $G_{E,M}^N(q^2) = G_{E,M}^N(0)$. 
However, UHECR–boosted C$\nu$B interactions involve significantly larger momentum transfers, and the $q$-dependence of nucleon form factors must be taken into account. 
Therefore, our cross section in Eq.~\eqref{eq:incoherent2}, which incorporates the full $q$-dependence of the nucleon form factors, is more general and directly applicable to UHECR–boosted C$\nu$B scattering.

\section{COSMIC RAY SOURCE DISTRIBUTIONS AND UHECR FLUX}
\label{subsec:source_param}

{\bf Star formation rate (SFR) source distribution.} The cosmic star formation rate density follows the empirical parameterization from \cite{Hopkins:2006bw}
\begin{equation}
\rho(z) = \frac{(a + b z)h}{1 + (z/c)^d} \ M_\odot \, \text{yr}^{-1} \, \text{Mpc}^{-3}\,,
\label{eq:SFR_density}
\end{equation}
where $h \equiv H_0 / (100 \, \mathrm{km \, s^{-1} \, Mpc^{-1}})$ is the dimensionless Hubble parameter, with $a = 0.0170$, $b = 0.13$, $c = 3.3$, and $d = 5.3$. The normalized source distribution function$f(z)$ is defined as
\begin{equation}
f(z) \equiv \frac{\rho(z)}{\rho(z_\mathrm{min})}\,,
\label{eq:norm_condition}
\end{equation}
where $\rho(z)$ represents the raw source density at redshift $z$, and $z_\mathrm{min} = 0$ serves as the baseline normalization redshift.

{\bf Quasistellar object (QSO) source distribution.} The logarithmic density evolution follows \cite{Wall:2004tg}:
\begin{equation}
\log \frac{\rho(z)}{\mathrm{Mpc}^{-3}} = -a_0 + a_1 z - a_2 z^2 + a_3 z^3 - a_4 z^4\,,
\label{eq:QSO_log}
\end{equation}
\begin{equation}
f(z) = \frac{\exp(\log\rho(z))}{\exp(\log\rho(z_\mathrm{min}))}\,,
\label{eq:QSO_norm}
\end{equation}
with coefficients $a_0 = 12.49$, $a_1 = 2.704$, $a_2 = 1.145$, $a_3 = 0.1796$, and $a_4 = 0.01019$.

{\bf Gamma-Ray burst (GRB) source distribution.} The GRB rate is modeled by scaling the star formation rate (SFR) density following \cite{Lan:2021uuf}:
\begin{equation}
\rho_{\text{GRB}}(z) = \kappa \rho(z) (1 + z)^\delta\,,
\label{eq:GRB}
\end{equation}
where $\rho(z)$ is the SFR density defined in Eq.~\eqref{eq:SFR_density}, $\kappa = 8.5$, and $\delta = 1.26$. The normalized density function $f(z)$ is then computed using Eq.~\eqref{eq:norm_condition}.

{\bf UHECR flux from PriNCe simulation.}
We simulate the UHECR flux at redshift $z$ produced by different CR source populations using the numerical propagation code PriNCe~\cite{Heinze:2019jou}. PriNCe solves the transport equations for UHECR nuclei in energy and redshift, including cosmological expansion, continuous energy losses due to photoproduction of electron-positron pairs, and photohadronic and photonuclear interactions.
The injected spectrum for each nuclear species $\mathcal{N}_i$ is parametrized as
\begin{equation}
J_{\mathcal{N}_i}(E_{\mathcal{N}_i})=
f_{\mathcal{N}_i}
\left(\frac{E_{\mathcal{N}_i}}{10^9\,\mathrm{GeV}}\right)^{-\gamma}
\times
\begin{cases}
1\,, & E_{\mathcal{N}_i} < Z_i R_{\max}\,, \\[4pt]
\exp\!\left(1-\dfrac{E_{\mathcal{N}_i}}{Z_i R_{\max}}\right)\!, &
E_{\mathcal{N}_i} > Z_i R_{\max}\,,
\end{cases}
\label{eq:PriNCe_injection}
\end{equation}
where $R_{\max}$ denotes the maximum rigidity, $\gamma$ is the spectral index,
and $f_{\mathcal{N}_i}$ are the normalized elemental abundances at injection.

The photon fields taken into account in the fit are the cosmic microwave background, scaling with redshift as usual, and the extragalactic background light as modeled by \cite{Gilmore2012}, which is also redshift  dependent. The photonuclear losses are described according to the TALYS model \cite{Koning2005}, while the photohadronic interactions are described by the SOPHIA code \cite{Mcke2000}. 

The parameters used in our PriNCe simulations are summarized in Table~\ref{tab:prince_parameters}.
The fitting parameters were determined in \cite{Sandrock:2025nzb} by a fit of PriNCe calculation results for a given source density to Pierre Auger Observatory results \cite{Veberic:2017hwu} on the total spectrum, depth of shower maximum $X_\text{max}$ and variance of the depth of the shower maximum $\sigma(X_\text{max})$, making use of the companion package \texttt{PriNCe-analysis-tools}\footnote{\url{https://github.com/joheinze/PriNCe-analysis-tools/}} developed by the authors of the PriNCe code.

\begin{table}[h]
\centering
\begin{tabular}{|c|c|c|c|}
\hline
 & SFR & QSO & GRB \\
\hline
$R_{\max}$ (GV) & $10^{9.25}$ & $10^{9.20}$ & $10^{9.25}$ \\
$\gamma$        & $-0.8$      & $-1.0$      & $-0.8$ \\
\hline
$f_{\rm H}$     & $1.0\times10^{-45}$ & $2.2\times10^{-45}$ & $1.7\times10^{-46}$ \\
$f_{\rm He}$    & $4.5\times10^{-46}$ & $7.7\times10^{-46}$ & $4.6\times10^{-47}$ \\
$f_{\rm N}$     & $6.2\times10^{-47}$ & $6.1\times10^{-47}$ & $5.5\times10^{-47}$ \\
$f_{\rm Si}$    & $4.1\times10^{-48}$ & $2.3\times10^{-48}$ & $3.9\times10^{-48}$ \\
$f_{\rm Fe}$    & $9.0\times10^{-50}$ & $6.6\times10^{-50}$ & $9.5\times10^{-50}$ \\
\hline
\end{tabular}
\caption{Fit parameters of the UHECR source models for SFR, QSO, and GRB distributions used in the PriNCe simulations~\cite{Sandrock:2025nzb}.}
\label{tab:prince_parameters}
\end{table}

{\bf Hillas cosmic ray energy spectrum.}
We adopt the Hillas parametrization to describe the present-day CR energy spectrum observed at Earth. This phenomenological model provides a standard description of the all-particle CR spectrum, and we use the parametrization introduced in Refs.~\cite{Hillas:2005cs, Gaisser:2013bla}.
It includes the standard three-population structure: the knee associated with Galactic supernova remnants, the ankle corresponding to the transition to extragalactic sources, and an intermediate region whose physical origin remains uncertain. The differential flux of a nucleus of type $\mathcal{N}_i$ at the present epoch can be written as~\cite{Hillas:2005cs, Gaisser:2013bla}
\begin{equation}
\left.\frac{d \phi_{\mathcal{N}_i}}{d E_{\mathcal{N}_i}}\right|_{z=0}
= \sum_{j=1}^{3}
a_{i,j}\,E_{\mathcal{N}_i}^{-\gamma_{i,j}-1}\,
\exp\!\left(-\frac{E_{\mathcal{N}_i}}{Z_i R_{c,j}}\right),
\label{eq:cr_flux}
\end{equation}
where $Z_i$ is the atomic number of the nucleus, $R_{c,j}$ is the cutoff rigidity of population $j$, 
and $a_{i,j}$ and $\gamma_{i,j}$ are obtained from fits to cosmic-ray observations.
Table~\ref{tab:hillas_params} summarizes the parameters for the five nuclear components (p, He, N, Si, Fe).

\begin{table}[h]
\centering
\begin{tabular}{|c|c|c|c|c|c|}
\hline
 & p & He & N & Si & Fe \\
\hline
Population~1: & 7860 & 3550 & 2200 & 1430 & 2120 \\
$R_c = 4 \, \text{PV}$ 
        & 1.66 & 1.58 & 1.63 & 1.67 & 1.63 \\
\hline
Population~2: & 20 & 20 & 13.4 & 13.4 & 13.4 \\
$R_c = 30 \, \text{PV}$ 
        & 1.4 & 1.4 & 1.4 & 1.4 & 1.4 \\
\hline
Population~3: & 1.7 & 1.7 & 1.14 & 1.14 & 1.14 \\
$R_c = 2 \, \text{EV}$ 
        & 1.4 & 1.4 & 1.4 & 1.4 & 1.4 \\
\hline
\end{tabular}
\caption{Hillas model parameters for different CR populations and nuclear species~\cite{Gaisser:2013bla}.}
\label{tab:hillas_params}
\end{table}

To match the Hillas model to current UHECR observation~\cite{Veberic:2017hwu}, 
we introduce an overall normalization factor $k$ to rescale the Hillas parameters~\cite{Gaisser:2013bla}. To test this normalization factor using the PAO data, we define the $\chi^2$ as~\cite{Sarmah:2025uzk,Sarmah:2024kek}
\begin{equation}
\label{chi2-hillas}
\chi^2(k) = \sum_i 
\frac{\bigl[J_{\mathrm{obs}}^{\,i} - k\,J_{\mathrm{Hillas}}^{\,i}\bigr]^2}{\sigma_i^2}\,,
\end{equation}
where $J_{\mathrm{obs}}^{\,i}$ denotes the UHECR flux values measured by PAO~\cite{Veberic:2017hwu},
$J_{\mathrm{Hillas}}^{\,i}$ is the corresponding flux predicted by the Hillas parametrization~\cite{Gaisser:2013bla}, and $\sigma_i$ represents the experimental uncertainty~\cite{Veberic:2017hwu}.
By minimizing the $\chi^2$ function with respect to $k$, we obtain the best-fit normalization factor $k_\text{best-fit} = 0.601$.
This normalization factor is adopted throughout our analysis whenever the Hillas model is used to describe the present-day UHECR spectrum.

Following the approach of Ref.~\cite{Herrera:2024upj}, the UHECR flux at redshift $z$ is obtained by rescaling the present-day CR spectrum using the source-distribution
function. It can be written as
\begin{equation}
\frac{d \phi_{\mathcal{N}_i}}{d E_{\mathcal{N}_i}}(z)
= f(z)\,
\left.\frac{d \phi_{\mathcal{N}_i}}{d E_{\mathcal{N}_i}}\right|_{z=0}\,,
\label{eq:hillas_redshift}
\end{equation}
where $f(z)$ denotes the CR source distribution function corresponding to the SFR, GRB, and QSO evolutions.
This rescaling procedure yields the UHECR flux at redshift $z$ without including UHECR propagation effects and is used for comparison with the PriNCe simulations.
Figure.~\ref{fig:cosmic_ray_flux} shows the resulting present-day CR spectrum, which
contains a larger fraction of heavy nuclei than the PriNCe result.

{\bf Energy loss of UHECR.}
Similar to the treatment of cosmic-ray energy losses induced by scattering with dark matter~\cite{Cappiello:2018hsu}, we estimate the energy loss of UHECR nuclei scattering with the C$\nu$B.
In each UHECR--C$\nu$B scattering, an energy $E_\nu$ is transferred from the UHECR to the relic neutrino, resulting in an energy loss of the UHECR.
In the continuous limit, the energy loss rate of a UHECR due to scattering can be written as
\begin{equation}
\frac{d\mathcal{Q}}{dt}
= \int_{0}^{E_{\nu}^{\max}} 
c\,\eta\,n_\nu\,E_{\nu}\,
\frac{d\sigma^{\nu \mathcal{N}_i}}{dE_{\nu}}\,dE_{\nu}\,,
\label{Energy loss of UHECR}
\end{equation}
where $\mathcal{Q}\equiv \Delta E_{\mathcal{N}_i}$ denotes the energy loss of a CR nucleus $\mathcal{N}_i$, $c$ is the speed of light, and the relation between cosmic time and redshift is given by $dt = dz \big[(1+z)\,H_0\sqrt{\Omega_{m}(1+z)^3+\Omega_{\Lambda}}\big]^{-1}$.
The upper integration limit $E_{\nu}^{\max}$ denotes the maximal neutrino energy kinematically allowed in the scattering process, as defined in Eq.~\eqref{eq:E-max}.

\begin{figure}[t!]
    \centering
    \includegraphics[width=0.49\textwidth]{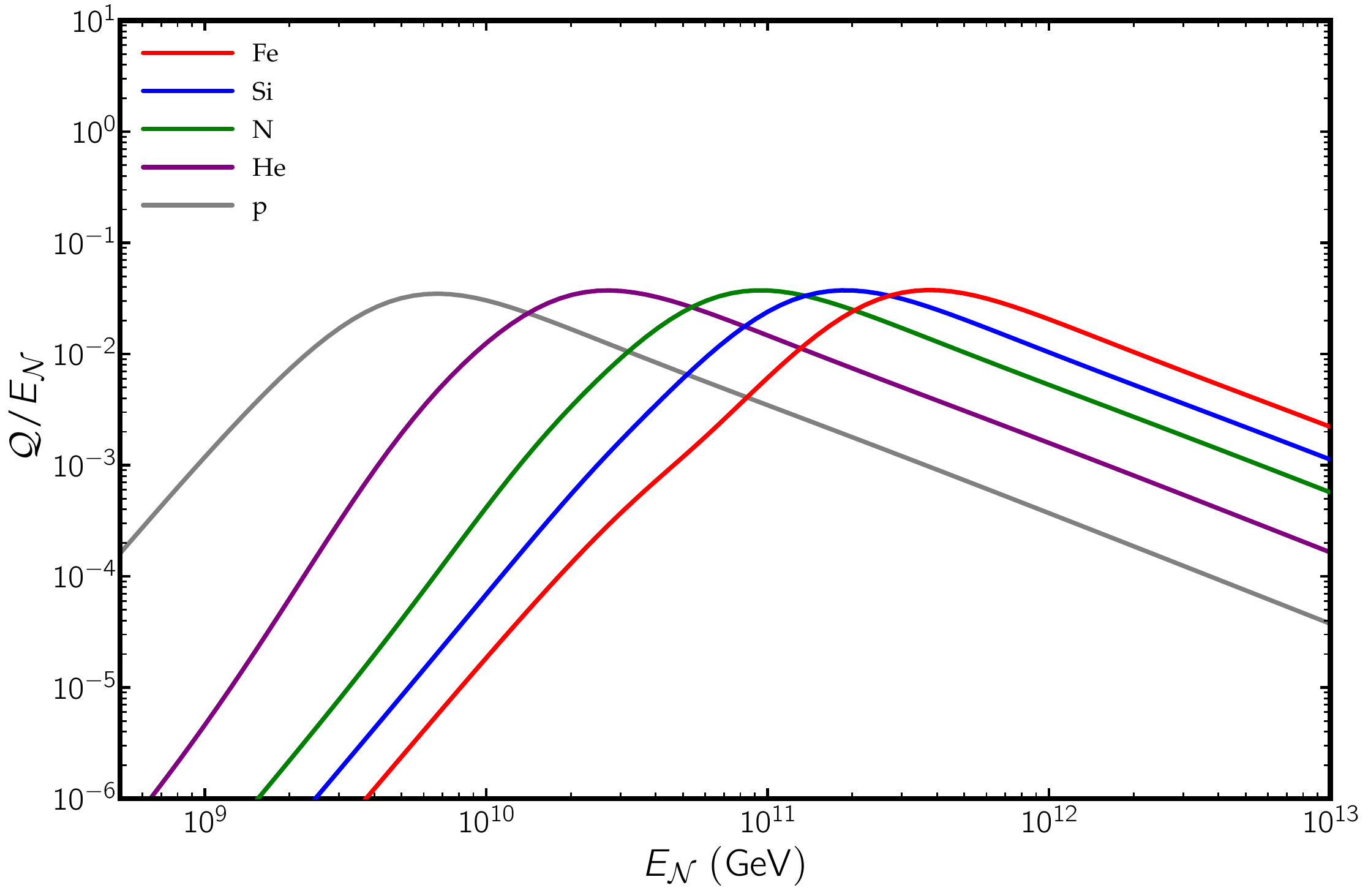}
    \caption{Energy-loss ratio $\mathcal{Q}/E_{\mathcal{N}_i}$ of CRs for protons, helium,nitrogen, silicon, and iron nuclei for $m_\nu = 0.1~\mathrm{eV}$ and $\eta = 10^{8}$. The redshift is integrated from 0 to 6.}
    \label{fig:Energy_loss_ratio}
\end{figure}

To estimate the largest energy loss of UHECR due to scattering with the C$\nu$B, we consider the longest possible propagation time that evolved from $z=6$ to $z=0$.
We set the relic neutrino mass to $m_\nu = 0.1\,\mathrm{eV}$ and the overdensity factor to $\eta = 10^{8}$. The impact of C$\nu$B scattering on UHECR propagation is quantified by the energy loss ratio $\mathcal{Q}/E_{\mathcal{N}_i}$.
In Fig.~\ref{fig:Energy_loss_ratio}, we show the energy loss ratio $\mathcal{Q}/E_{\mathcal{N}_i}$ for different UHECR nuclear components. The ratio increases with increasing CR energy at low energies and decreases at higher energies.
The initial rise is caused by the rapid growth of the scattering cross section with increasing CR energy.
At sufficiently high energies, the cross section is suppressed by the nuclear form factor and no longer increases, leading to a decrease of the energy-loss
ratio. This behavior is consistent with the energy dependence of the scattering cross section shown in Fig.~\ref{fig:cross_section} in the main text. At the peak of the energy loss ratio, the value remains far below unity, indicating that C$\nu$B scattering has a negligible effect on UHECR propagation.
We also notice that if the C$\nu$B overdensity were extremely large, the energy loss of UHECR could become non-negligible. A detailed analysis of how to incorporate the C$\nu$B scattering into UHECR propagation is left for future work.

\begin{figure}[t!]
    \centering
    \includegraphics[width=0.49\textwidth]{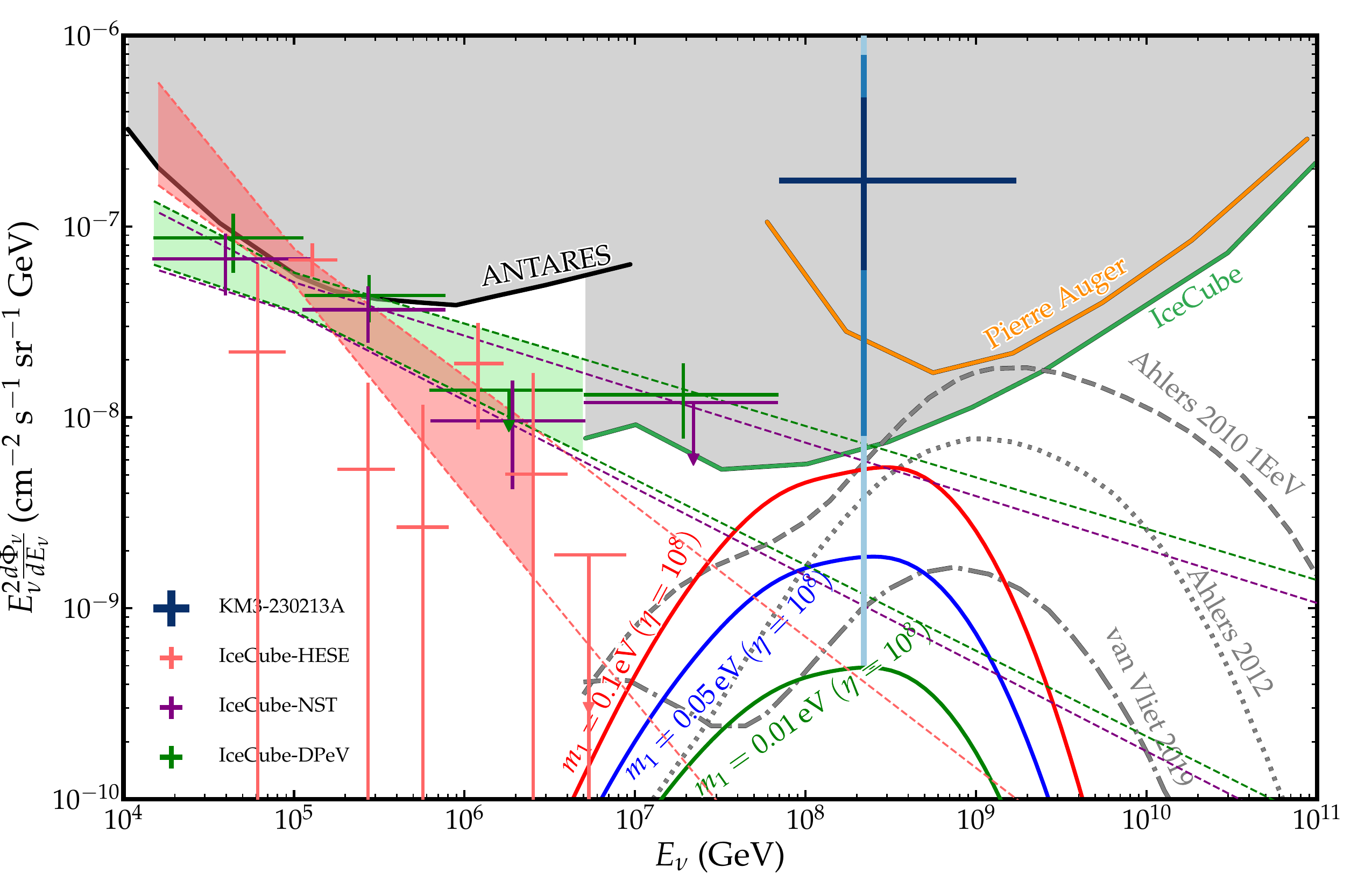}
    \includegraphics[width=0.49\textwidth]{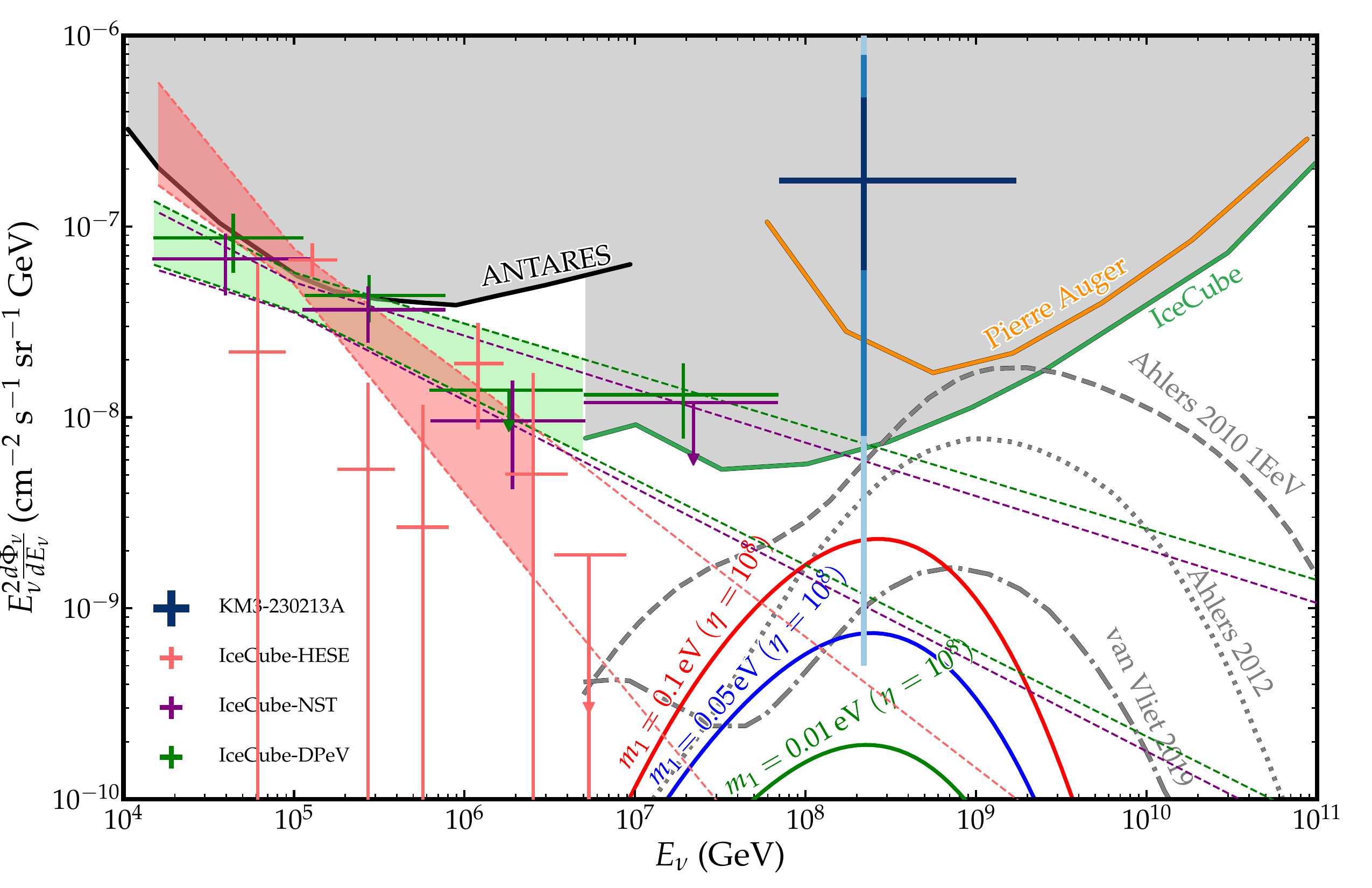}
    \includegraphics[width=0.49\textwidth]{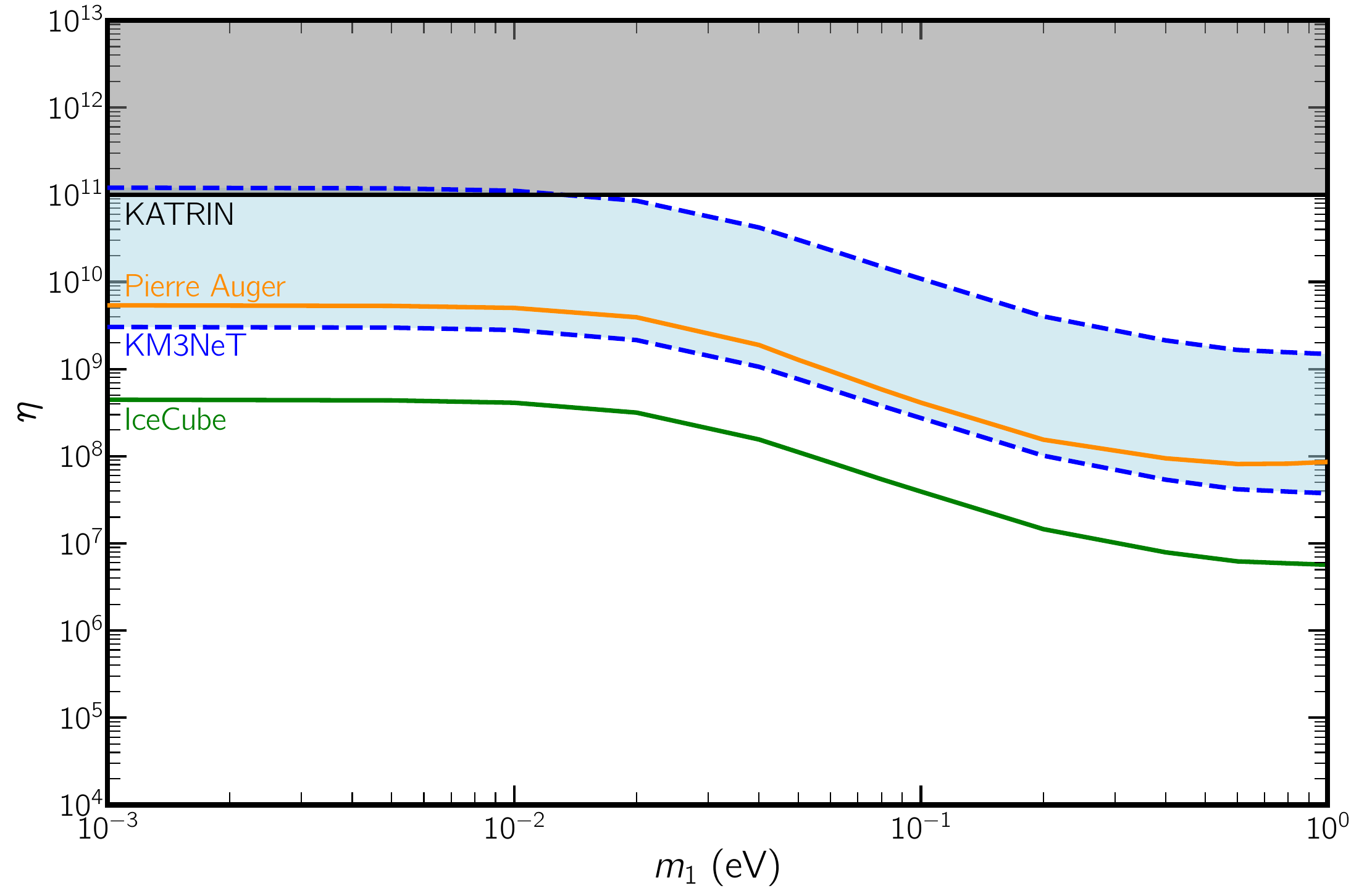}
    \includegraphics[width=0.49\textwidth]{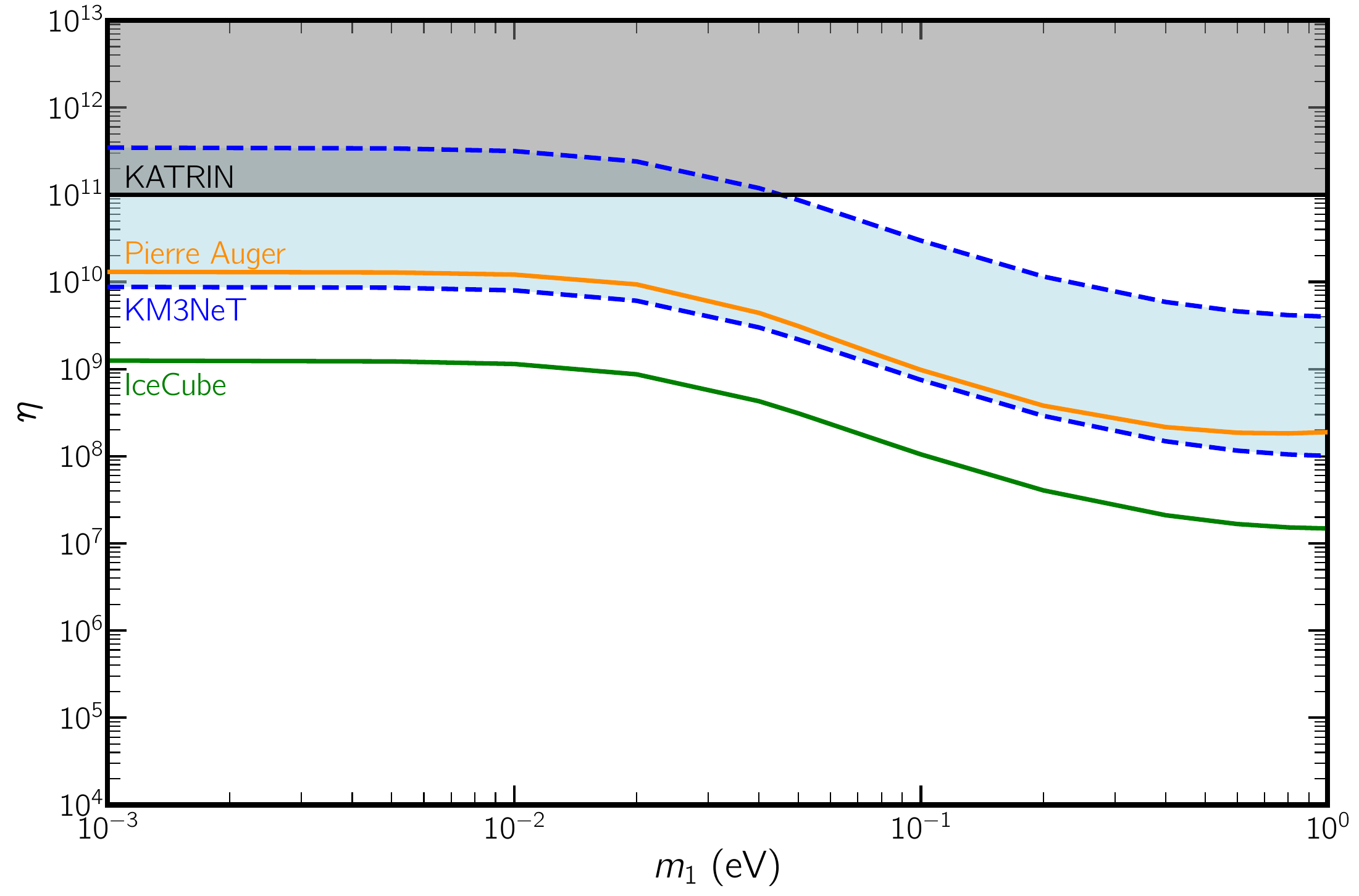}
    \caption{Same as Fig.~4 in the main text, except for using the GRB (left panels) and QSO (right panels) model for the CR source distribution.}
    \label{fig:GRB-QSO}
\end{figure}

{\bf Results for GRB and QSO source distributions.}
Figure.~\ref{fig:GRB-QSO} illustrates the boosted C$\nu$B flux and the 90\% CL constraints on the overdensity parameter $\eta$ for different CR source distributions, including GRB (left panels) and QSO (right panels) models. For $m_1=0.1 \, \text{eV}$, IC constrains $\eta < 3.9 \times 10^7$, while PAO constrains $\eta < 4.1 \times 10^8$. With a decrease in the lightest neutrino mass to $m_1 = 0.01 \, \text{eV}$, IC constrains $\eta < 4.1 \times 10^8$, and PAO constrains $\eta < 5.0 \times 10^9$. Among the source distribution models considered, the GRB source distribution yields the most stringent constraints on the overdensity factor $\eta$, while the QSO source distribution leads to the weakest constraints. The constraints from the SFR source distribution lie between those obtained from the GRB and QSO source distributions.

\section{COMPARISON OF THE ELASTIC NEUTRINO-PROTON CROSS SECTION}
\begin{figure}[t!]
    \centering
    \includegraphics[width=0.49\textwidth]{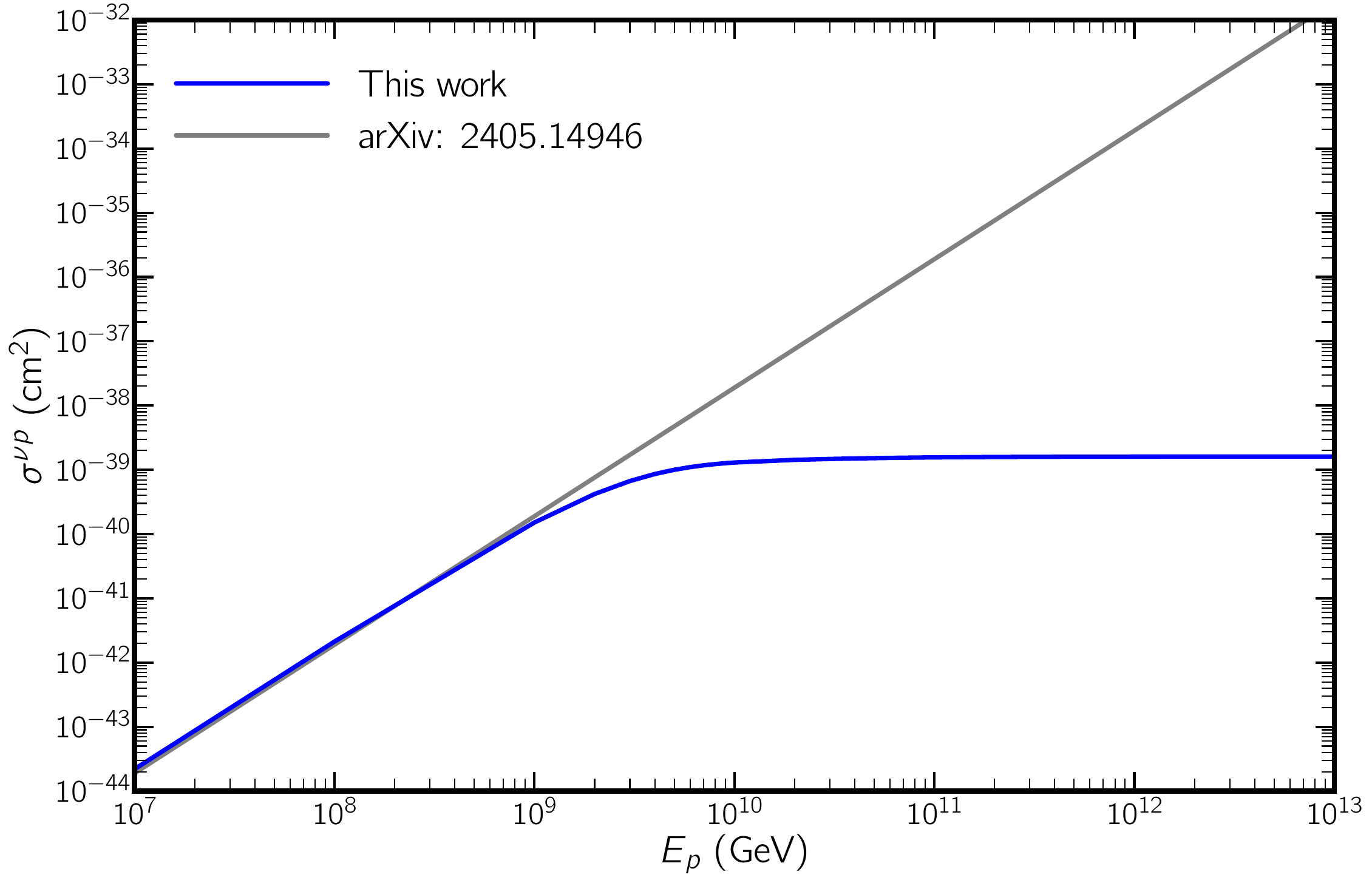}
    \caption{The neutral current neutrino-proton cross section as a function of the proton energy. The blue line indicates the energy-dependent cross section adopted in this work, and the gray line shows the one used in Ref.~\cite{Herrera:2024upj}.}
    \label{fig:Compare}
\end{figure}

Here, we compare the neutrino-proton elastic scattering cross section with the analysis of Reference.~\cite{Herrera:2024upj}, which result in an overestimation of the boosted C$\nu$B flux and overly stringent constraints on the overdensity parameter $\eta$. The key issue lies in their treatment of the differential scattering cross section for UHECRs interacting with the C$\nu$B. Ref.~\cite{Herrera:2024upj} use $\sigma^{\nu p} / E_\nu^{\text{max}}$ as an approximation for $d \sigma^{\nu p}/d E_\nu$  [see Eq.~(1) in Ref.~\cite{Herrera:2024upj}]. However, as shown in Fig.~\ref{fig:Compare}, the $\sigma^{\nu p}$ used in Ref.~\cite{Herrera:2024upj} only agrees with our results at low energies, and it is several orders of magnitude larger than ours at high energies.
The main reason is that the $\sigma^{\nu p}$ used in Ref.~\cite{Herrera:2024upj} has
neglected the energy dependence of the nucleon form factor in  $d \sigma^{\nu p}/d E_\nu$. In particular, $d \sigma^{\nu p}/d E_\nu$ approach zero at high energies. This oversimplification of cross section will lead to an inflated C$\nu$B flux prediction. Similar issues have also been discussed in the context of boosted dark matter~\cite{Bardhan:2022bdg}.

\section{IMPACT OF THE LIGHTEST NEUTRINO MASS AND THE INVERTED MASS ORDERING}
\begin{figure}[t!]
    \centering
    \includegraphics[width=0.49\textwidth]{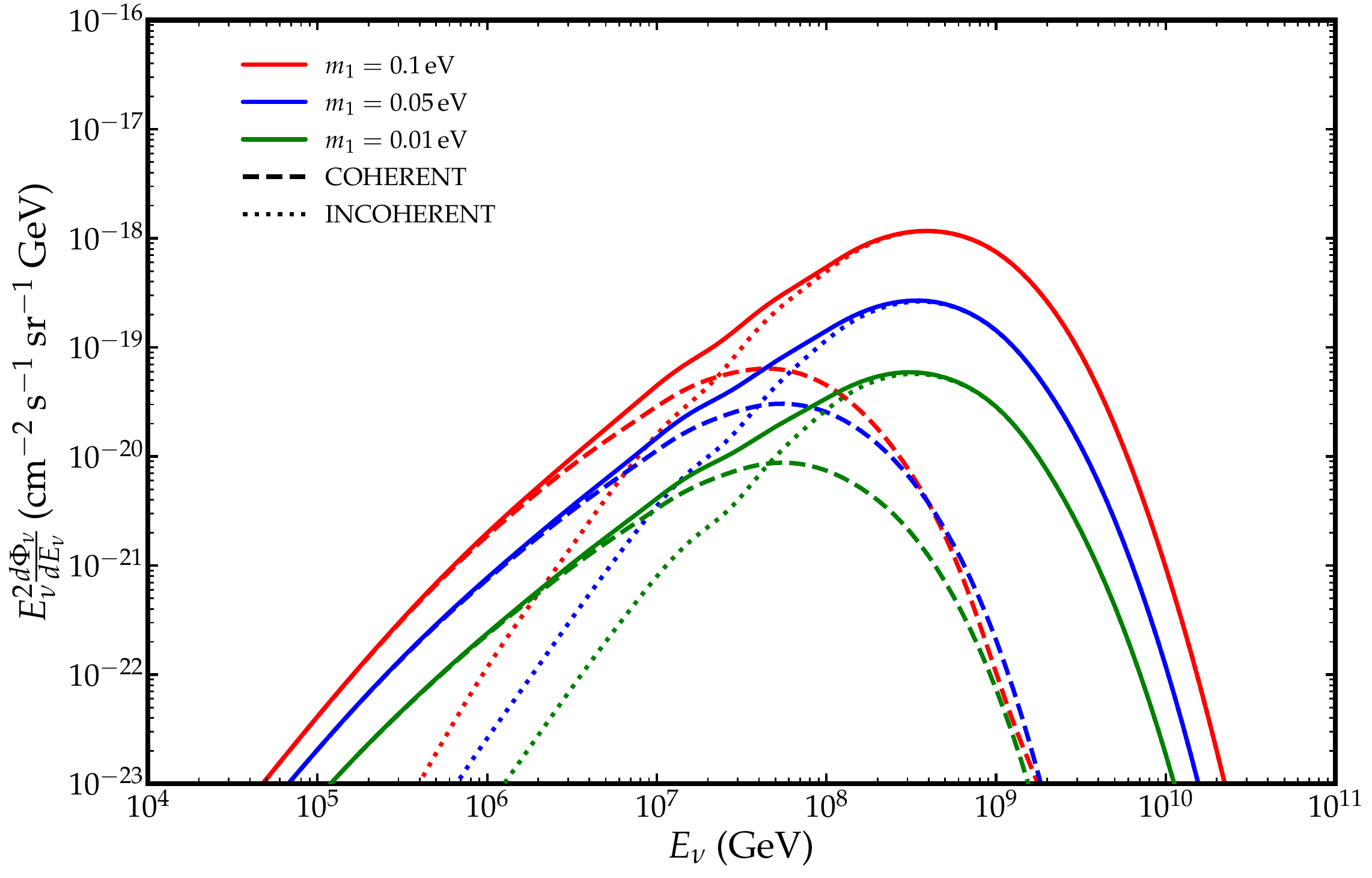}
    \caption{The total boosted C$\nu$B flux at Earth for different lightest neutrino masses $m_1$. Solid lines show the total flux from all nuclei (excluding protons) in red ($m_1 = 0.1 \, \text{eV}$), blue ($m_1 = 0.05 \, \text{eV}$), and green ($m_1 = 0.01 \, \text{eV}$), with dashed and dotted lines correspond to the coherent and incoherent contributions, respectively. Here, we use the SFR model for CR source evolution with $\eta = 1$.}
    \label{fig:flux_supp}
\end{figure}

\begin{figure}[t!]
    \centering
    \includegraphics[width=0.49\textwidth]{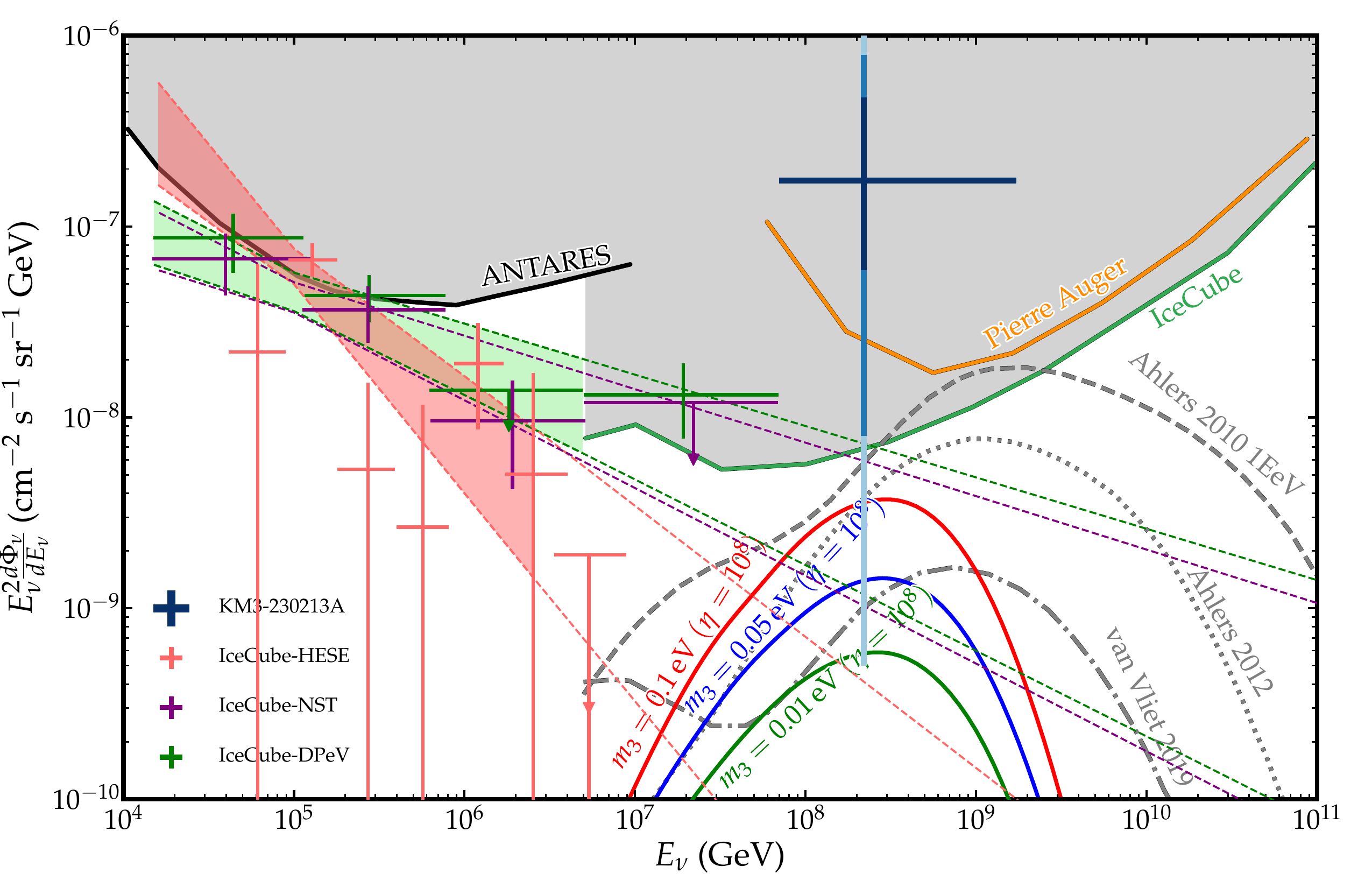}
    \includegraphics[width=0.49\textwidth]{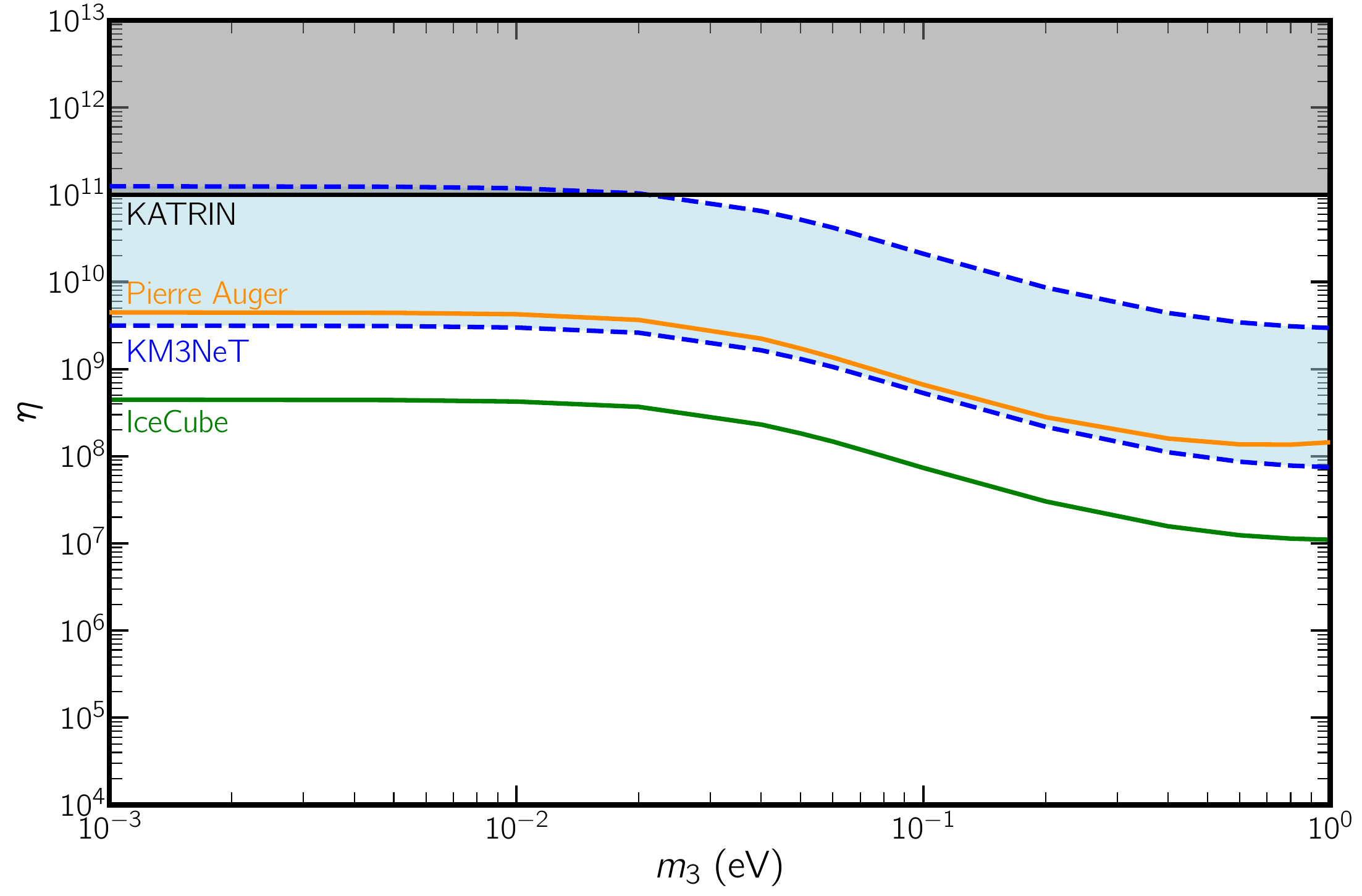}
    \caption{Same as Fig.~4 in the main text, except for the case of inverted neutrino mass ordering.}
    \label{fig:Inverted}
\end{figure}

In the case of the normal neutrino mass ordering, we calculate the contributions of coherent and incoherent scattering to the boosted C$\nu$B flux for different lightest neutrino masses, as shown in Fig.~\ref{fig:flux_supp}, considering all nuclei except for protons. As the lightest neutrino mass $m_1$ decreases, the boosted C$\nu$B flux decreases. However, for $m_1 < 0.01 \, \text{eV}$, the total flux no longer changes, as the dominant contributions arise from the heavier masses $m_2$ and $m_3$, which are less affected by the reduction in $m_1$. Furthermore, since the momentum transfer $q^2 = 2 m_\nu E_\nu$ depends on the neutrino mass, a smaller $m_1$ reduces $q^2$ at a given energy $E_\nu$, thereby extending the energy range where coherent scattering dominates to a higher energy.

To investigate the impact of the mass ordering on the boosted C$\nu$B flux, we plot Fig.~\ref{fig:Inverted}, which adopts the same setup as Fig.~4 in the main text except for an inverted mass ordering with $\Delta m_{23}^2 = 2.486 \times 10^{-3} \, \text{eV}^2$ and $\Delta m_{21}^2 = 7.42 \times 10^{-5} \, \text{eV}^2$~\cite{Esteban:2020cvm}. In the left panel, the flux for $m_3 = 0.1 \, \text{eV}$ and $m_3 = 0.05 \, \text{eV}$ shows negligible differences compared to the normal ordering, indicating that the boosted flux remains relatively insensitive to these values for the lightest mass. However, for $m_3 = 0.01 \, \text{eV}$, the neutrino flux increases by approximately 50\% compared to the normal ordering. Similarly, in the right panel, this feature is also observed for $m_3 < 0.01 \, \text{eV}$.


\bibliography{references}

\end{document}